\documentclass[%
groupedaddress,
 amsmath,amssymb,
 aps,
prl,
twocolumn,
floatfix,
notitlepage
]{revtex4-1}

\usepackage{graphicx}
\usepackage{color}
\usepackage{dcolumn}
\usepackage{bm}
\usepackage{subfigure}
\usepackage{tabularx}
\makeatletter
\renewcommand{\thesubfigure}{\alph{subfigure}}
\renewcommand{\@thesubfigure}{\thesubfigure)\hskip\subfiglabelskip}
\makeatother
\usepackage{hyperref}

\def \x{\bm{x}}

\def \S{\bm{S}}
\def \s{\bm{s}}

\def \k{\bm{k}}

\def \q{\bm{q}}
\def \r{\bm{r}}

\def \pt{\partial}

\def \im{\text{Im}}
\def \re{\text{Re}}

\def \beq{\begin{equation}}
\def \eeq{\end{equation}}
\def \bspt{\begin{split}}
\def \espt{\end{split}}
\def \bef{\begin{figure}}
\def \enf{\end{figure}}
\def \tr{\text{Tr}}
\def \bpm{\begin{pmatrix}}
\def \epm{\end{pmatrix}}
\newcommand{\abs}[1]{\lvert#1\rvert}
\newcommand{\braket}[2]{\mbox{$\langle #1 | #2 \rangle$}}
\newcommand{\meanvalue}[3]{\mbox{$\langle #1 | #2 | #3 \rangle$}}
\newcommand{\proj}[2]{\mbox{$|#1\rangle \!\langle #2 |$}}
\newcommand{\ev}[1]{\mbox{$\langle #1 \rangle$}}
\newcommand{\bra}[1]{\mbox{$\langle #1 |$}}
\newcommand{\ket}[1]{\mbox{$| #1 \rangle$}}

\begin{document}


\title{
Anomalous Hall effect and quantum criticality in geometrically frustrated heavy fermion metals
}

\author{Wenxin Ding$^{1,2,\dagger}$}
\author{Sarah Grefe$^{3,2,\dagger}$}
 \author{Silke Paschen$^4$}
  \author{Qimiao Si$^2$}%
  \affiliation{%
    $^1$School of Physics and Optoelectronics Engineering, Anhui University, Hefei, Anhui Province, 230601, China\\
    $^2$Department of Physics \& Astronomy, Rice University, Houston, Texas 77005, USA\\
    $^3$Department of Physics \& Astronomy, California State University, Long Beach, California 90840, USA\\
    $^4$Institute of Solid State Physics, Vienna University of
 Technology, Wiedner Hauptstra{\ss}e 8-10, 1040 Vienna, Austria
 }%

 \begin{abstract}
Studies on the heavy-fermion pyrochlore iridate (Pr$_2$Ir$_2$O$_7$) point to the role of time-reversal-symmetry breaking in geometrically frustrated Kondo lattices.  Here we address the effect of Kondo coupling and chiral spin liquids in a $J_1-J_2$ model on a square lattice and a model on a kagom\'{e} lattice via a large-$N$ method, based on a fermionic representation of the spin operators, and consider a new mechanism for anomalous Hall effect for the chiral phases.  We calculate the anomalous Hall response for the chiral states of both the Kondo destroyed and Kondo screened phases.  Across the quantum critical point, the anomalous Hall coefficient jumps when a sudden reconstruction of Fermi surfaces occurs.  We discuss the implications of our results for the heavy-fermion pyrochlore iridate and propose an interface structure based on Kondo insulators to explore such effects further.
\end{abstract}

\maketitle
\begin{figure}[t!]
  \centering
\includegraphics[width=0.9\columnwidth]{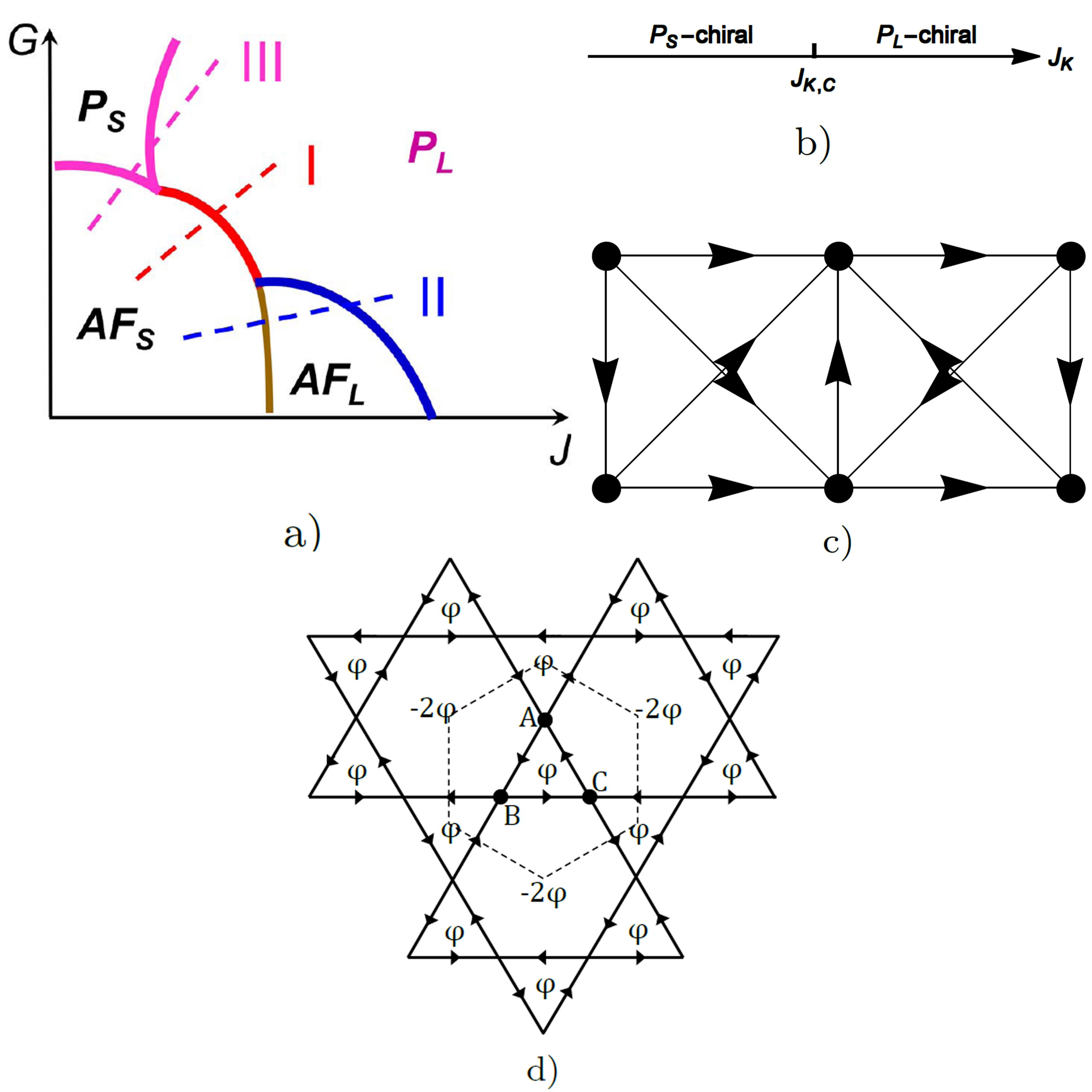}
  \caption{(Color online)
(a) The  global phase diagram of Kondo lattice systems \cite{Si2010a};
(b)
In the highly frustrated regime (large, fixed $G$),
$J_K$
tunes
through a Kondo-destruction QCP (at $J_{K,c}$)
from a Kondo-destroyed chiral spin liquid ($P_{S,chiral}$) to a Kondo-screened phase ($P_{L,chiral}$).
The
$\chi$ fields of the square lattice are shown in the $\pi$-flux state (without the diagonal bonds) and the CSL state (c),
and in
the CSL state on the kagom\'{e} lattice (d):
the arrows denote the sign of gauge field $A_{ij}$, and $\phi$ is the flux through a triangle.
}
\label{fig:fig1}
\end{figure}

Heavy-fermion metals are prototypical systems to study quantum criticality
\cite{Coleman2005,Hu-Natphys2024,Kirchner2020}.
The simplest model to describe these systems
is a Kondo lattice, which comprises a lattice of local moments and a band of
conduction electrons. The local moments are coupled to each other by the
Ruderman-Kittel-Kasuya-Yosida (RKKY) interactions and are simultaneously
connected to a band of conduction electrons through an antiferromagnetic (AF)
Kondo exchange interaction ($J_K$).  In recent years,  it has been
realized that the effect of geometrical frustration is a potentially fruitful
but little-explored
frontier. From a theoretical perspective, geometrical frustration
enhances $G$, the degree of quantum fluctuations in the magnetism of the
local-moment component, and
a $J_K - G$ phase diagram at zero temperature has been advanced
\cite{Si2010a,Coleman2010}.
Figure~(\ref{fig:fig1}a)
illustrates the proposed global phase diagram \cite{Si2010a},
which applies the notion of Kondo-destruction \cite{Si-Nat2001} to the parameter space that incorporates the frustration and related quantum fluctuation effects.
From a materials
perspective, there is a growing effort in
studying frustrated Kondo-lattice
compounds
\cite{KimAronson2013,Mun2013,Fritsch2014,Tokiwa2015,Nakatsuji2006,SiPaschen2013,Zhao2019,Pas-Si2021}

The pyrochlore
heavy-fermion
 system
 Pr$_2$Ir$_2$O$_7$
 is one such example. Both the measured
magnetic susceptibility and specific heat \cite{Nakatsuji2006} suggest
the presence of Kondo coupling between the Ir $d$-electrons and the local $f$-moments of Pr.
No magnetic order is found down to very low temperatures,
suggesting that the
$f$-moments of Pr develop a quantum spin liquid
(QSL)
ground state \cite{Nakatsuji2006}. In addition,  experiments found a
sizeable zero-field anomalous Hall effect (AHE)
for magnetic field applied along the [111] direction
\cite{Machida2007,Machida2010}, revealing a spontaneous
time-reversal-symmetry-breaking (TRSB) state.

This system
is of considerable theoretical interest \cite{Chen2012a, Flint, Lee2013b,Moon2013,Savary2014,Udagawa2013,Kalitsov2009}.
With a few exceptions \cite{Rau2013}, the role of the Kondo effect has not been discussed in this context,
and neither has its relationship with
the observed quantum criticality.
Yet, the
observation of a large entropy and a divergent Gr\"{u}neissen ratio \cite{Tokiwa2014}
clearly point to the importance of the Kondo coupling and the role of a proximate heavy-fermion quantum critical point (QCP).
In the case of AF heavy-fermion systems,
the normal Hall effect has been successfully used to probe the evolution of the Fermi surface across the QCP and, thereby, the nature of quantum
criticality \cite{Paschen2004,Kirchner2020,Pas-Si2021}.
Given that the AHE is also
intrinsically a Fermi surface property (other than contributions from fully occupied bands)~\cite{Haldane2004},
we are motivated to address whether it can serve as a diagnostic tool for the QCP in the present setting.
In addition to
elucidating the
AHE,
studying this issue
promises
to
bring about the much-needed new understanding of
quantum phases and their transitions in
geometrically-frustrated heavy-fermion
metals
\cite{Pas-Si2021}.
Given the complexity of the
three-dimensional pyrochlore lattice,
we will gain insights from
related but simpler models.

In this Letter, we study both the frustrated $J_1-J_2$ quantum Heisenberg model on a
square lattice and the $J_1$ only model on the kagom\'{e} lattice with a Kondo coupling to conduction electrons.
For the square lattice, we consider
the regime of strong frustration where a chiral spin liquid (CSL) phase~\cite{Wen1989}
becomes energetically competitive in
a
large-$N$ limit
(based on a Schwinger fermion representation of the spin operators; see below).
The kagom\'{e} lattice, representing
a layer perpendicular to the [111] direction of the pyrochlore lattice,
is a two-dimensional network of corner-sharing triangles
[Fig.~(\ref{fig:fig1}d)]
 with a strong geometrical frustration.
A CSL phase is
found
in a spin-$\frac{1}{2}$ model
on the kagom\'{e} lattice~\cite{kagomeCSL}.
Using the large-$N$ limit
\cite{Lhuillier2015}, we will also study the CSL physics on this lattice.
We develop the method to
calculate the AHE in both a Kondo-destroyed ($P_S$) and a Kondo-screened ($P_L$)
paramagnetic phase.
We show that each phase may have a sizable AHE. Moreover,
across a QCP,
the AHE jumps when the Fermi surface suddenly reconstructs.

{\it
Frustrated Kondo-lattice models~}
We study the following Hamiltonian:
\begin{equation}
  \label{eq:H}
  \begin{split}
    H & =  H_{f}  + H_{d,0}  + H_{K} \quad .
  \end{split}
\end{equation}
Here $H_{f}$ describes a Heisenberg model.
For the square lattice case, $H_f$ includes both $J_1$ and $J_2$ couplings between
the nearest neighbors ($nn$, $\ev{}$) and next-nearest neighbors ($nnn$,$\ev{\ev{}}$).
We focus on the maximally frustrated case of
$J_2/J_1 = 1/2$.
For the kagom\'{e} case,  the lattice is geometrically frustrated and
it suffices for
$H_f$
to only contain
the $nn$ term.
For both models with $H_f$ alone,
CSL states
appear in the large-$N$ limit \cite{Wen1989, Affleck1988}.

The local moments are coupled to a band of conduction electrons, described by
$H_{d,0} = - \sum_{ij, \alpha} (t_{ij} d^\dagger_{i\alpha} d_{j\alpha} +h.c.) $,
through an AF Kondo
 interaction $J_K$, specified by
$H_K = J_K \sum_i \s_i\cdot\S_i$.
Here, $\s_i = \sum_{\alpha,\beta}\frac{1}{2}d^\dagger_{i\alpha} \bm{\sigma}_{\alpha\beta} d_{i\beta}$
is
the spin of the conduction electrons, with $\bm{\sigma}$
describing the Pauli matrices.
We
take $t_{\ev{ij}} = t = 1$ as the energy unit.

We  use the Schwinger fermion representation for the $f$-moments
$\bm{S}_i = \sum_{\alpha,\beta} \frac{1}{2} f^{\dagger}_{i\alpha} \bm{\sigma}_{\alpha\beta} f_{i\beta}$,
with the constraint $\sum_{\alpha} f^{\dagger}_{i \alpha} f_{i \alpha} = 1$,
so that $H_{f} = \sum_{\alpha,\beta,ij} \frac{J_{ij}}{2}  f^{\dagger}_{i \alpha} f_{i \beta}
 f^{\dagger}_{j\beta} f_{j\alpha} - \frac{J_{ij}}{4} n_{i \alpha} n_{j \beta}$.
 In
 the
 large-$N$ approach \cite{Affleck1988},
the spin index $\alpha = 1, 2, \dots, N$,
 and the constraint is enforced by a Lagrangian multiplier $\lambda_i$.
The Heisenberg and Kondo terms are decoupled by a Hubbard-Stratonovich (HS) transformation.
The large-$N$ limit leads to
\begin{equation}\label{eq:MF}
    H_{eff} = H_{QSL} + H_{d,0} + H_{K,eff} + E_c \quad ,
\end{equation}
with $H_{QSL} = - \sum_{ij,\alpha} \frac{J_{ij}}{2} (\chi_{ji} f^\dagger_{i\alpha}  f_{j\alpha}  + h.c.)
- \sum_{i,\alpha} \lambda_i
(f^\dagger_{i \alpha} f_{i\alpha} - 1/2)  $, $H_{K,eff} = - \sum_{i,\alpha} \frac{J_K}{2}
(\pi_i d^\dagger_{i \alpha} f_{i \alpha} + h.c.) $,
and $E_c = \sum_{ij} N J_{ij} \abs{\chi_{ij}}^2/2 + \sum_{i} N J_K \abs{\pi_i}^2/4 $.
The HS fields are defined as $\chi_{ij} = \sum_{\alpha}\ev{f_{i \alpha}^\dagger f_{j \alpha}}$
and $\pi_i = \sum_{\alpha}\ev{f^\dagger_{i \alpha}d_{i \alpha}}$.
Both can be decomposed into amplitudes and phases: $\chi_{ij} =  \rho_{ij} e^{iA_{ij}}$, $\pi_i = \rho_{K,i} e^{iA_{K,i}}$.
The Kondo parameter $\pi_i$
can be taken to be real,
with its phase absorbed
into the
 field $\lambda_i$,
 i.e.\ $\pi_i \rightarrow \rho_{K,i}$.

By minimizing the total energy of $H_{eff}$ in Eq.~(\ref{eq:MF}), we obtain the phase diagrams
containing the chiral states, in which
$J_K$
tunes the system from
the  $P_S$ to $P_L$ phases
(see Supplemental Material~\cite{supp}).
Across a second-order Kondo-destroyed $P_{S,chiral}$ to $P_{L,chiral}$ quantum phase transition,
Fig.~(\ref{fig:fig1}b),
we consider a power-law form
for the Kondo hybridization amplitude:
\begin{equation}
  \label{eq:rho_k-function}
  \rho_K(J_K) = \rho_r \Big(\frac{J_{K}-J_{K,c}}{J_{K}}\Big)^{1/2} \quad ,
\end{equation}
for $J_K > J_{K,c}$ and $\rho_K=0$, for $J_K < J_{K,c}$.
We  take $J_{K,c}$ as the value where the $P_{L,chiral}$ state
becomes
energetically competitive
and $\rho_r$ to be the saturation value of $\rho_K$;
both values are adopted from the self-consistent calculation
for a given set of $(n_d, J_1)$ ~\cite{supp}.

{\it Mechanism of the AHE
-
the Kondo destroyed
$P_S$ phase.~}
In the Kondo-destroyed $P_S$ phase, the static
hybridization amplitude vanishes,  $\ev{\rho_{K,i}} = 0$.  However, we
show that
there are TRSB terms in the effective interactions among the conduction
electrons, which are mediated by the spinons via Kondo couplings. Such terms
yield a zero-field
AHE.

We will single out the TRSB terms.
The TRSB order parameter of the CSL is the spin chirality,
\begin{equation}
  \hat{E}_{ijk} = \S_i \cdot (\S_j \times \S_k) \quad ,
\end{equation}
where the indices $\{i,j,k\}$
mark the three sites of an elementary triangle of the lattice. In the CSL state,
$E_{ijk} = \ev{\hat{E}_{ijk}} = 2 i (P_{ijk } - P_{ikj})$, where $P_{ijk} = \chi_{ij}\chi_{jk} \chi_{ki}$.
On symmetry grounds, we expect $E_{ijk}$ to be coupled to the composite chiral operator of the conduction electrons,
$\s_i \cdot (\s_j \times \s_k)$. With this guidance,
we obtain the coupling from integrating out the $f$-fermions and expanding in powers of $J_K$; this can be represented by
triangular diagrams (Supplemental Material~\cite{supp}), similar to what is used in deriving a chiral current.
We find
\begin{equation}\label{eq:Hcc}
  \begin{split}
    & H_{\text{chiral}} = \sum \frac{J_K^3}{3!} \underbrace{(\s_i\cdot \S_i)  (\s_j\cdot \S_j)
    (\s_k\cdot \S_k)}_{\triangle \text{-loop  contraction}} \\
    & =  \frac{ J_K^3}{2 \times 3!} E_{ijk} \s_i \cdot (\s_j \times \s_k) \quad .
  \end{split}
\end{equation}

In the kagom\'{e} case, the hexagons can also possess non-trivial fluxes.
However, we can restrict the effective TRSB coupling for the conduction electrons to the lowest order in $ J_K$, which corresponds to considering only the fluxes of the triangles.

The chiral interactions in $H_{\text{chiral}} $ have a six-fermion form.
We can decouple it by introducing a novel HS transformation
that involves triangular diagrams described in the Supplemental Material\cite{supp}.
We end up with an effective bilinear theory:
\begin{equation}
  \label{eq:eff-chiral-H}
  H_{d} = H_{d,0} + H_{d,1}
\end{equation}
with
\begin{eqnarray}
&H_{d,1} = \sum_{ij} (g \phi_j^* \phi_i d^\dagger_i d_j+ \phi^*_i G^{-1}_{\phi,ij} \phi_j + h.c. ) \quad .
\end{eqnarray}
Hence, the $\phi$-fields are constrained by the condition that, if they are integrated out,
we obtain the same chiral interaction terms at $\mathcal{O}(g^3)$ by computing
the same triangle diagrams.
We then replace $\phi_j^* \phi_i$ by its expectation value $G_{\phi,ij}$ and arrive at
\begin{equation}
  H_{d,1} \rightarrow \sum_{ij} (g G_{\phi, ij} d^\dagger_i d_j  + h.c.) \quad .
\end{equation}
 It turns out that $G_{\phi,ij} = e^{- i A_{ij}}$,
 and $g$ can be identified as $g = J_K ( \abs{E_{ijk}} / 2)^{1/3}$.
 Because the bosonic Gaussian integral has a minus sign relative to its fermionic counterpart,
 $G_{\phi,ij}$ carries the opposite flux pattern to produce the same $H_{\text{chiral}}$
 when we integrate out the $\phi$-fields.
 Physically, the flux (or chirality) pattern has the opposite sign to that of the CSL state,
 so that the antiferromagnetic Kondo coupling will lower the ground state energy.
This effective Hamiltonian is adequate for qualitatively describing the AHE physics
of our original Hamiltonian.
Other non-chiral effective interactions
would only renormalize the Fermi liquid parameters of the $d$-electrons
for the $P_S$ phase.
We can then use the Streda formula \cite{Streda,Nagaosa2010}
 to compute the AHE coefficient $\sigma_{xy}$:
 The involved quantities are
 the current operator of the conduction electrons $v_a({\bf k}) = \pt_a H_{d}({\bf k})$,
the Berry curvature $\mathcal{F}^{xy}_n({\bf k}) $,
and the Fermi function $f(\epsilon_n({\bf k}))$
(Supplemental Material\cite{supp}).

{\it Mechanism of the
AHE - the Kondo screened $P_L$ phase.~}
In the $P_L$ phase, the Kondo order parameter $\rho_{K,i}$ acquires a non-zero expectation value
$\rho_K = \ev{\rho_{K,i}}$.
There should still be an incoherent piece
of the slave boson fields:
 $\rho_{K,i} = \rho_K + \pi'_{i}$.
Moreover, we focus on the case where the chiral order survives in the $P_L$ phase.
By considering the same triangular diagrams now mediated by the incoherent part $\pi'_i$,
the fluctuations of the Kondo order parameter still mediate chiral interactions similarly
as in the $P_S$
phase, but with a reduced weight. However,
there is no spectral sum rule for the $\pi'_i$s
to obtain this reduced weight readily.
 In the Supplemental Material~\cite{supp}, we use
a slave rotor approach for the periodic Anderson model to determine this factor.
The effective Hamiltonian of the $d$-electrons becomes
\begin{equation}
H_{d} = H_{d,0} + [1 - (4J_K/U) \rho_K^2 ] H_{d,1} \quad ,
\end{equation}
where $U$ is the onsite Hubbard repulsion. We fix $U = 2W$, {\it i.e.}\ twice
the $d$-electron's
bandwidth throughout the calculations.
Keeping only the $\rho_K$ part of $H_K$ leads to
 the following
effective Hamiltonian:
\begin{equation}\label{eq:mean-field-H}
  H_{P_L} = \Psi^\dagger
  \begin{pmatrix}
    H_{\text{CSL}} & - J_K \rho_K \mathcal{I}  \\
    - J_K \rho_K \mathcal{I} & H_{d}
  \end{pmatrix}
  \Psi \quad ,
\end{equation}
where $\mathcal{I}$ is an identity matrix, and  $\Psi^\dagger =    (f^\dagger,   d^\dagger).$
We have dropped the spin index,
as there are no longer spin-flip terms.
The Hamiltonian $H_{P_L}$ is smoothly connected with $H_d$ at the QCP.
We then compute $\sigma_{xy}$ from
the Streda formula Eq.~(S-22),
noting that
the  current operators
remains the same, i.e. \ $v_a({\bf k}) = \pt_a H_{d}({\bf k})$.

{\it AHE and its evolution across the Kondo-destruction QCP.~}
For the square lattice, we focus on the $\pi$-flux and the CSL states which are known to be closely competing in the pure $J_1-J_2$ Heisenberg models. In the large-N calculation based on Eq.~(\ref{eq:mean-field-H}), we find that both states can be stabilized in the presence of Kondo screening. The $P_L$-CSL state emerges as the ground state first, but the $P_L-\pi$-flux state eventually takes over at larger $J_K/J_1$ (see Fig. (S5) and related Supplemental Material for details).

For the $\pi$-flux phase, $H_{QSL} = H_{\pi -flux}$ is given by $A_{\r_i, \r_i+\hat{x}}
 = \pi/2$, $A_{\r_i, \r_i + \hat{y}} = - (-1)^{x_i}\pi/2 $,  $\rho_{\r_i, \r_i + \hat{x} + \hat{y}} = 0$,
 where $\r_i = (x_i, y_i)$, $\hat{x}$ ($\hat{y}$) is the unit vector along the $x(y)$-axis.
 For the CSL Hamiltonian, $H_{QSL} = H_{CSL}$ is derived from $H_{\pi -flux}$
 with $\rho_{\r_i, \r_i + \hat{x} + \hat{y}} \neq 0$, $A_{\r_i, \r_i + \hat{x} + \hat{y}} =
 A_{\r_i+\hat{y}, \r_i + \hat{x} } = (-1)^{x_i}\pi/2$,
 as illustrated
 in Fig.~(\ref{fig:fig1}c).

In the kagom\'{e} lattice, any state with triangle flux $\phi\neq0,\pi$ breaks TRS. Here, we choose $\phi=-\frac{\pi}{2}$ such that the hexagon flux of $-2\phi=\pi$ preserves TRS. The $(-\frac{\pi}{2},\pi)$ spinon flux state has three well-separated bands; the middle flat band is exactly at the Fermi energy, and the Chern numbers are $C_{lower}=-1$,  $C_{middle}=0$, $C_{upper}=+1$~\cite{Nagaosa2010}. These band structures can considered as the usual, no-flux kagom\'{e} bands inverted by the fluxes. The phase structure of the corresponding $\chi_{ij}$ fields is plotted in Fig.~(\ref{fig:fig1}d).

\begin{figure}[t!]
  \centering
  \subfigure[]{\label{sfig:AHE-1}\includegraphics[width=0.45\columnwidth]{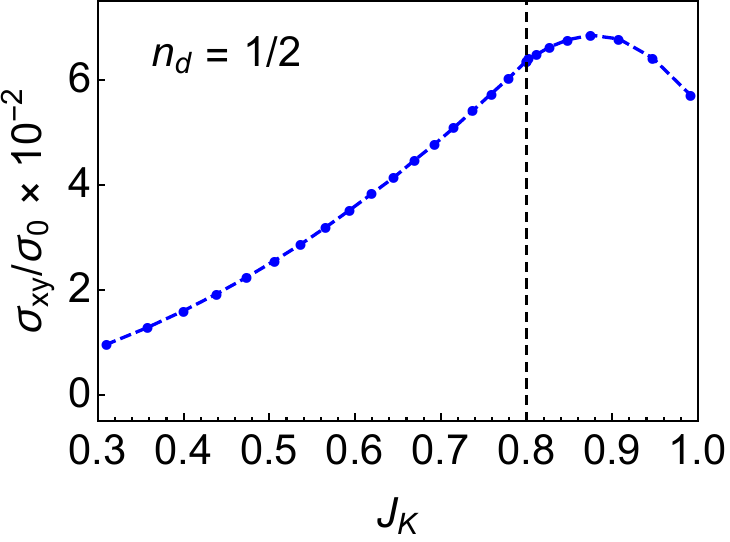}}
  \subfigure[]{\label{sfig:kagAHE}\includegraphics[width=0.45\columnwidth]{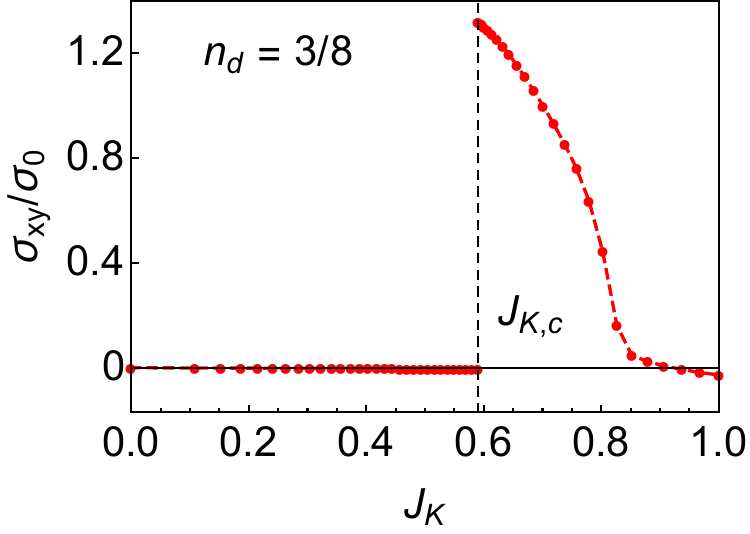}}
  \caption{(Color online)  Zero field anomalous Hall conductance ($\sigma_{xy}$),
  normalized by the quantum conductance $\sigma_0 = e^2/h$,
  for $J_1 =t= 1,\ J_2/J_1=1/2, \ n_d = 0.5$ on a square lattice
  (a) and for $J=t,\ n_d=3/8$ on a Kagom\'{e} lattice (b).
 }
\label{fig:AHE}
\end{figure}

The zero-field anomalous Hall conductivity
$\sigma_{xy}$ of the $J_1-J_2-J_K$ model is shown in Fig.~(\ref{sfig:AHE-1}) for a representative parameters
$n_d = 0.5,\ J_1 =  t$ and
that  of the kagom\'{e} lattice model in Fig~(\ref{sfig:kagAHE}) for $J=t,\ n_d = 3/8$.
Across the QCP, $\sigma_{xy}$ is found \emph{continuous} for the square lattice, but \emph{jumps discontinuously} for the kagom\'{e} lattice. The amplitudes of $\sigma_{xy}$ are similar in the $P_S$ regimes, $\sim 10^{-2} \sigma_0$. But $\sigma_{xy}$ is enhanced by two orders of magnitude in the
$P_L$ regime of the
kagom\'{e} case.

\begin{figure}[t!]
  \centering
  \subfigure[]{\label{sfig:BerryCurvature3}\includegraphics[width=0.45\columnwidth]{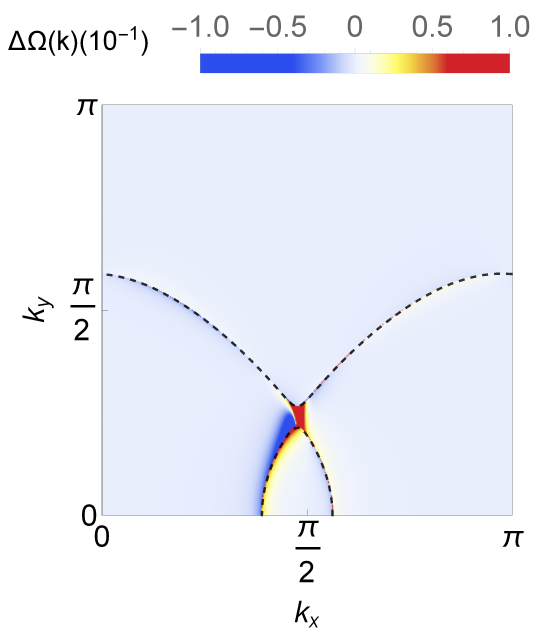}}
  \subfigure[]{\label{sfig:BCFS_delta}\includegraphics[width=0.45\columnwidth]{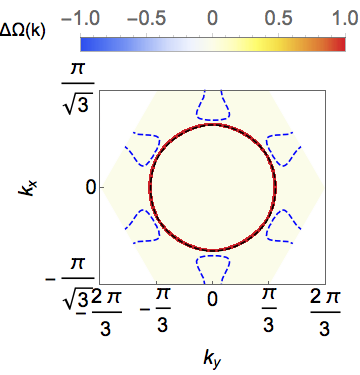}}
  \caption{(Color online) Fermi surfaces (dashed curves) and the difference
  in the band-summed Berry curvature distribution $\Delta \Omega ({\bf k})$ between the $P_S$ phase and the $P_L$ phase
  (color map) of the square lattice model (\ref{sfig:BerryCurvature3}) and the kagom\'{e} lattice model (\ref{sfig:BCFS_delta}).
  }\label{fig:kagome}
\end{figure}

In order to understand the different behaviors, we show the Fermi surfaces (dashed lines) and the difference of band-summed Berry curvature $\Delta \Omega ({\bf k})$ (color map)  between the $P_S$ phase and the $P_L$ phase right across the QCP in Fig. (\ref{sfig:BerryCurvature3}) for the square lattice and (\ref{sfig:BCFS_delta}) for the kagom\'{e} lattice (the actual $\Omega ({\bf k})$ is shown in Supplemental Material).
Here $\Delta \Omega ({\bf k})= \Omega_{P_S} ({\bf k}) - \Omega_{P_L} ({\bf k})$ and $\Omega ({\bf k}) = \sum_n \mathcal{F}^{xy}_n({\bf k}) f(\epsilon_n({\bf k}))$.
We find the Fermi surfaces remain continuous for the square lattice model.
Both Fermi surfaces of the $P_S$ and $P_L$ phases are the black dashed line.
However, for the kagom\'{e} case, the Fermi surfaces show a jump.
The Fermi surface of the $P_S$ phase is the black dashed circle in the middle of the BZ
which overlaps with the red, singular part of $\Delta \Omega ({\bf k})$.
Those of the $P_L$ phase are the blue dashed-line pockets at the edge of the BZ.
These results reflect
the number of sites per unit cell
and the gapped/gapless nature of the spinon spectrum.
However, $\Delta \Omega ({\bf k})$ is singular and concentrates near Fermi surfaces in both cases.
This is because the onset of Kondo hybridization,
which acts as a topological mass term in the large-$N$ theory,
in general singularly reconstructs the wavefunctions regardless of whether the Fermi surfaces jump or not.

To reconcile the notions of the singular wavefunction (or Berry curvature) with the continuous AHE,
we note that $\sigma_{xy}$ is intrinsically a Fermi surface property~\cite{Haldane2004}
(apart from the contributions of fully occupied bands).
We can analytically show the following by
computing the diagonal Berry's connection in the $\rho_K \rightarrow 0$ limit~\cite{supp}.
When the Fermi surfaces {evolve continuously across the QCP}, $\sigma_{xy}$
must be continuous; here,
the projected wavefunctions of the $d$-electron are continuous, and so are the Berry connections.
By contrast, when the Fermi surface jumps, the projected wavefunctions
completely reconstruct
due to the existence of two non-commuting topological ``masses": the Kondo screening and a non-zero jump of the
spinon Lagrangian multiplier $\lambda$.

{\it Discussion}.
Energetic considerations\cite{Coleman1989a} show that the Kondo coupling favors
 gapless states (Supplemental Material~\cite{supp}), since the formation of a Kondo singlet generally needs to overcome the spinon gap, if any,
 lowering
 the energy by an amount $\propto J_K$.
For the pyrochlore lattice, the CSL state in the large-$N$ limit is
gapless~\cite{Burnell2009}, and is
thus expected to
have a
similar sequence of quantum phase transitions involving the chiral state.
The gapless nature
raises the prospect of
a sudden reconstruction of the Fermi surface across a Kondo-destruction QCP
in the pyrochlore case and, by extension,
a jump in the zero-field AHE,
especially for a magnetic field along the [111] direction.

We expect the jump of the zero field AHE, $\sigma_{xy}$, to be robust against weak disorder.
The AHE effect considered here is intrinsic, i.e., determined by the quasi-particle band structure.
Scattering from weak non-magnetic impurities only
yields
a small (linear in disorder) correction
~\cite{Sinitsyn2007}. Moreover, the Fermi-surface jump across a Kondo-destruction QCP
has been evidenced to be robust against weak disorder~\cite{Gegenwart2008,SiPaschen2013}.
Thus, our results can be tested in Pr$_2$Ir$_2$O$_7$, once a control parameter is identified to tune across
the implicated zero-field QCP~\cite{Tokiwa2014}.
From Ref.~\onlinecite{Ohtsuki2019},
strain can potentially serve as such a
tuning parameter.

We note that the anomalous Hall conductance from the mechanism advanced here
is quite large.
Experiments in Pr$_2$Ir$_2$O$_7$ \cite{Machida2010}
find a large
sheet $\sigma_{xy}$
reaching about 0.7\% of
$\sigma_0 \equiv {\rm e}^2/h$, a value which readily arise in
our mechanism (Fig.~\ref{sfig:AHE-1}).

We have emphasized the role of the Kondo effect and its critical destruction.
Future work should
incorporate {\it ab initio} features, not only on the directional dependence in the pyrochlore lattice
but also
the effect of the {\it ab initio} electronic band structure
and
the non-Kramers nature
of the ground-state crystal-field level of the Pr
ions~\cite{Chandra2013b,Rau2013}. Although Ref. \cite{Rau2013} studied
a realistic model, the mechanism of AHE in the Kondo destroyed phase is not captured, and $\sigma_{xy}^{AHE}$ only starts to grow from zero at the QCP. The new mechanism we consider here provides a possible interpretation to the experimental observation that $\sigma_{xy}^{AHE}$ is most singular at the QCP.

Furthermore, we have derived our conclusions in
geometrically frustrated Kondo systems
and demonstrated the robustness of our results by connecting them with
the evolution of the Fermi surfaces. Thus, we expect our results to remain
qualitatively valid
in the more realistic settings.
For Pr$_2$Ir$_2$O$_7$,
this is so
given the substantial evidence for the role of the Kondo coupling
such as
the large entropy observed in the pertinent low-temperature regime \cite{Tokiwa2014}.
It may also be instructive to explore related effects in other $f$-electron systems with geometrical frustration,
such as UCu$_5$ under ambient conditions\cite{Ueland2012} and when suitably tuned through a QCP.
%
%
Recently, the
proximity to the
Kondo-destruction QCP
we predict here for
Pr$_2$Ir$_2$O$_7$ is confirmed experimentally\cite{Aynajian2020}. In this STM measurement, regions of heavy Fermi liquid are interweaved with a non-magnetic metallic phase with Kondo-destruction, forming spatial nanoscale patterns consistent with being in proximity to a critical point.

While the current work emphasizes the link between AHE and the evolution of Fermi surfaces due to Kondo physics, other topology-related components, such as conduction band topology, $k$-dependence of Kondo coupling, etc., are not taken into account. It is known that such components can lead to
topological Kondo lattice models that realize such topological states as Weyl-Kondo semimetals
\cite{Lai2018,Chen-Natphys2022,Dzsaber2017,Dzsaber2021}.
In addition, $d$-electron-based systems on frustrated lattices, through the notion of compact molecular orbitals, can realize topological Kondo lattice models and the associated states
\cite{chen2023Metallic,Chen-emergent-2024,HuSciAdv2023}.
For these states,
Berry-curvature-induced Hall effect
is also an important characteristic;
thus, we expect our work here will provide new insights
into those systems.

We close by proposing an engineered Kondo-insulator interface as a model
material for the frustrated Kondo lattice Hamiltonian.
The
motivation for the proposed setting comes
from
advances in the molecular beam epitaxy (MBE) of Kondo
systems
\cite{Shishido2010,Goh2012,Prochaska2020}.
As a promising candidate material, we suggest the golden
phase of SmS ($g$-SmS).  In bulk samples,
this phase is stable under pressures
between about 0.65\,GPa \cite{Maple1971}
 and 2\,GPa \cite{Haga2004}. As MBE
thin-film, the phase might be stabilized by lattice mismatch with an
appropriate substrate. $g$-SmS crystallizes in a face-centered-cubic (fcc)
structure of rock-salt (NaCl) type. A lattice plane is shown in
Fig.~(S6). $g$-SmS shows characteristics of a Kondo insulating state in
transport \cite{Wac94.1,Haga2004}, thermodynamics \cite{Matsubayashi2007}, and point
contact spectroscopy \cite{Wac94.1}. From thermal expansion and heat capacity
measurements, the energy gap was estimated to be 90\,K on the low-pressure side
of the $g$-SmS phase \cite{Matsubayashi2007}. At temperatures low compared to this scale,
the proposed lattice plane could then serve as a setting to realize the frustrated
$J_1-J_2$ Kondo lattice and study the anomalous Hall effect.

{\it Note added:} Since our manuscript was posted, chiral heavy fermion phase and its associated Kondo-destruction transition have also been discussed in the context of moir\'{e} structures of transition metal dichalcogenides \cite{Kumer2022,Guerci2023}.

We acknowledge useful discussions with P. Gegenwart, S. Nakatsuji
and P. Goswami.
Work at Rice University has primarily been supported by the National Science Foundation
under Grant No. DMR-2220603 (model conceptualization, W.D. and S.E.G.), by
the Air Force Office of Scientific Research under Grant No.
FA9550-21-1-0356 (model calculations, W.D. and S.E.G.),
by the Robert A. Welch Foundation Grant No. C-1411 and the Vannevar Bush Faculty Fellowship ONR-VB N00014-23-1-2870 (conceptualization, Q.S.). Work at Anhui University was supported by the National Key R\&D Program of the MOST of China under Grant No. 2022YFA1602603 (W.D.).

$\dagger$ These two authors contributed equally to this work.

%

\clearpage
\pagebreak
\setcounter{figure}{0}
\makeatletter
\renewcommand{\thefigure}{S\@arabic\c@figure}
\setcounter{equation}{0} \makeatletter
\renewcommand \theequation{S-\@arabic\c@equation}
\renewcommand \thetable{S\@arabic\c@table}

\begin{widetext}
  \begin{center}
    \textbf{\large Supplemental Materials}
  \end{center}
\end{widetext}

\subsection{Chiral interaction from perturbative expansion}

\begin{figure}
  \centering
  \includegraphics[width=8cm]{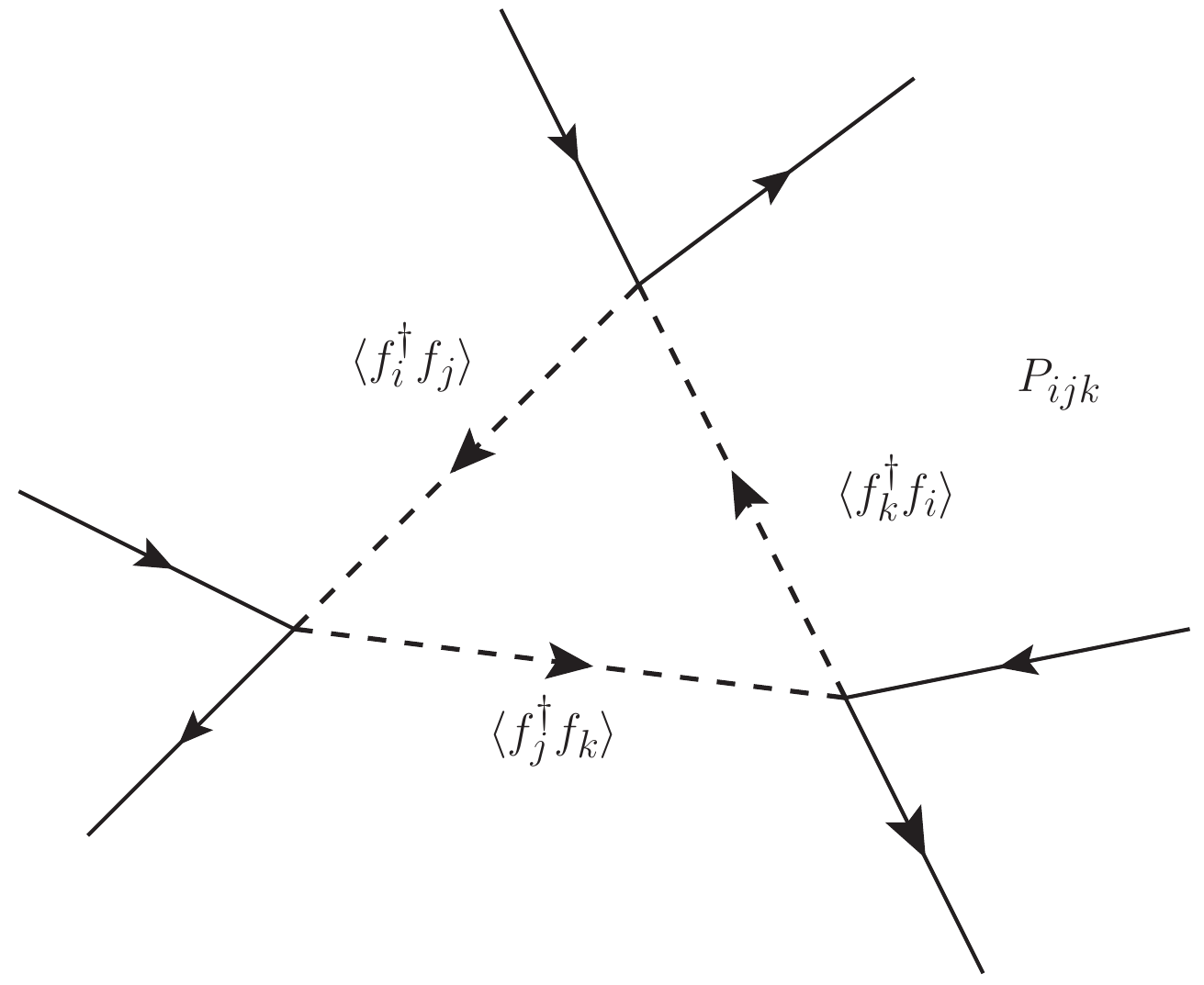}
  \caption{Feynman diagrams of the triangle-loop contractions. The solid lines are the propagators for the conduction electrons,
  and the dashed lines for the spinons.}
  \label{fig:chiral-int}
\end{figure}

{\it $H_{\text{chiral}}$ of the Kondo destroyed $P_S$ Phase}.
To obtain Eq. (5) of the main text, we integrate out the $f$-spinons from Eqs. (1,2) of the main text
in the Kondo destroyed phase
using the standard Feynman diagram procedure. Guided by
symmetry analysis,
we only need to consider the
third-order term $1/3! (\s_i\cdot \S_i) (\s_j\cdot \S_j) (\s_k\cdot \S_k) $.
The effective chiral electronic interaction $H_{cc}$ is obtained by contracting the spinons in triangle-loop
diagrams as shown Fig.~(\ref{fig:chiral-int}).

Since the CSL is gapped, it is sufficient to restrict to the most local three-site loops, i.e., the triangle within a unit cell,
at equal time only. Then we can obtain $H_{\text{chiral}}$ as follows:
\begin{equation}\label{eq:eff-chi-chi}
  \begin{split}
   &  (\s_i\cdot \S_i) (\s_j\cdot \S_j) (\s_k\cdot \S_k)  = \sum_{a,b,c} s_i^a  s_j^b  s_k^c \\
   & \times \frac{1}{8}(f^\dagger_{i\alpha_i} \sigma^a_{\alpha_i\beta_i}
   f_{i\beta_i} f^\dagger_{j\alpha_j} \sigma^b_{\alpha_j\beta_j}
   f_{j\beta_j} f^\dagger_{k\alpha_k} \sigma^c_{\alpha_k\beta_k} f_{k\beta_k})\\
&\underrightarrow{\triangle}  \sum_{a,b,c} s_i^a  s_j^b  s_k^c  (\ev{f^\dagger_{i\alpha_i}  f_{j\beta_j}}
\ev{f^\dagger_{j\alpha_j}  f_{k\beta_k}} \ev{f^\dagger_{k\alpha_k}  f_{i\beta_i}} \\
&+ \ev{f^\dagger_{i\alpha_i}  f_{k\beta_k}}\ev{f^\dagger_{k\alpha_k}  f_{j\beta_j}} \ev{f^\dagger_{j\alpha_j}
f
_{i\beta_i}} ) \sigma^a_{\alpha_i\beta_i} \sigma^b_{\alpha_j\beta_j} \sigma^c_{\alpha_k\beta_k}\\
& = E_{ijk} \s_i \cdot (\s_j \times \s_k)/2,
  \end{split}
\end{equation}
where $\underrightarrow{\triangle}$ denotes
the triangular-loop contraction.
The contraction is approximated by equal-time correlators, so $\ev{f^\dagger_{i \alpha_i}  f_{j \beta_j}}
= \delta_{\alpha_i,\beta_j} \chi_{ij}$.
To isolate the chiral processes, we can discard the density-density interactions.
As a result, the second line of Eq. (\ref{eq:eff-chi-chi}) can be written as
\begin{equation}
  \label{eq:Hcc}
  \begin{split}
  & H_{\text{chiral}} \sim \sum_{\alpha_l,\alpha_j,\alpha_k} (P_{ljk} d^\dagger_{l\alpha_k} d_{l\alpha_l}
   d^\dagger_{j\alpha_l} d_{j\alpha_j} d^\dagger_{k\alpha_j} d_{k \alpha_k} \\
  & + P_{lkj} d^\dagger_{l\alpha_j} d_{l\alpha_l} d^\dagger_{k\alpha_l} d_{k\alpha_k} d^\dagger_{j\alpha_k} d_{j \alpha_j}).
  \end{split}
\end{equation}

{\it $H_{\text{chiral}}$ of the Kondo Screened
$P_L$ Phase}.
To obtain the spectral weight of the incoherent terms in the Kondo screened phase,
we use the slave rotor theory\cite{Florens2004} to tackle the $f$-fermion Hubbard model.
As we shall briefly discuss below, the Kondo transition {\it is} the Mott transition for $f$-fermions in periodic
Anderson model (PAM)\cite{Florens2002}, and is realized when the rotor fields are condensed.
The condensation density describes the coherent charge degrees of freedom that would contribute to transport.

The PAM Hamiltonian is
\begin{equation}
\label{eq:periodic-Anderson-model}
  H_{\text{PAM}} = H^{(f')}_{\text{Hubbard}} + H_0^{(d)} + V \sum_{i,\sigma} (f'^\dagger_{i,\sigma} d_{i, \sigma}
  + f'_{i, \sigma} d^\dagger_{i, \sigma}),
\end{equation}
where
\begin{equation}
  \label{eq:f-Hubbard-model}
  H^{(f')}_{\text{Hubbard}} = -\sum_{ij,\sigma} t_{ij} f'^\dagger_{i,\sigma} f'_{j,\sigma} + U \sum_{i,\sigma,\sigma'}
  n^{(f')}_{i,\sigma} n^{(f')}_{i,\sigma'}
\end{equation}
is the usual half-filled Hubbard model, and
\begin{equation}
  \label{eq:d-free-electron}
  H_0^{(d)} = -\sum_{ij,\sigma} t_{ij} d^\dagger_{i,\sigma} d_{j,\sigma}
\end{equation}
describes the free $d$-band electrons.

First, we use the slave rotor formalism to treat the Hubbard model part by letting $f'_i \rightarrow
 f_i e^{- i\theta_i}$
\begin{equation}
  H^{(f')} \rightarrow \frac{U}{2} \sum_{i,\sigma} \hat{L}_{i,\sigma}^2 - \sum_{ij,\sigma}( t_{ij}
   f^\dagger_{i,\sigma} f_{j,\sigma} e^{i(\theta_i - \theta_j)} + h.c.).
\end{equation}
The corresponding Lagrangian is
\begin{equation}
  \begin{split}
    & S_H = \int d\tau \sum_{i,\sigma} f^\dagger_{i,\sigma} \pt_\tau
    f_{i,\sigma} + \frac{(\pt_\tau \theta) ^2}{2 U} +\\
    & \sum_{\ev{ij},\sigma} ( t_{\ev{ij}} f^\dagger_{i,\sigma}
    f_{j,\sigma} e^{i(\theta_i - \theta_j)} + h.c.),
  \end{split}
\end{equation}
here the kinetic energy of the rotors $\frac{U}{2} \sum_{i,\sigma} \hat{L}_{i,\sigma}^2$ is replaced
by its conjugate variables $\hat{L}_{i,\sigma} = (\pt_\tau \theta + i h)/U$.

Let $e^{i\theta_i}  = X_i$, so that $X_i$s subject to the constraint $\abs{X_i}^2 = 1$ on average
(using Lagrangian multiplier). Using $\pt_\tau \theta_i = \frac{1}{i} X_i^* \pt_\tau X_i$, we have
\begin{equation}
    S_H  = S^0_X + S_f^0 +  \sum_{\ev{ij},\sigma} ( t_{\ev{ij}} f^\dagger_{i,\sigma} f_{j,\sigma} X_i X^*_j + h.c.).
\end{equation}
with $S^0_f = \int d\tau \sum_{i,\sigma} f^\dagger_{i,\sigma} \pt_\tau  f_{i,\sigma} $, and $S_X^0
= \sum_i (\frac{\abs{\pt_\tau X_i}^2}{2 U} + \lambda_i (\abs{X_i}^2 - 1))$.
The exchange term is also expressed in terms of slave rotors
\begin{equation}
  \label{eq:exchange-H}
  H_{\text{exc}} = V \sum_{i, \sigma} (f^\dagger_{i, \sigma} d_{i, \sigma} X_i + h.c.).
\end{equation}

In the large-$U$-small-$V$ limit, the system is in $P_S$ phase. We can integrate out the rotor fields\cite{Ding2014},
and recover both the Heisenberg-$J$ interaction, as well as the Kondo coupling
\begin{equation}
H_K = J_K \sum_i \S_f(i) \cdot \s_d(i),
\end{equation}
where $J_K = 4 V^2/U$.

Within the slave rotor approach, the onset of Kondo screening is described by the condensation
of the $X$-field: $X_i \rightarrow X^0_i + X'_i$. The exchange term becomes
\begin{equation}
  \label{eq:rotor-hyb}
  H_{exc} =  \sum_{i, \sigma} ( V X^0_i f^\dagger_{i, \sigma} d_{i, \sigma}
  + V X'_if^\dagger_{i, \sigma} d_{i, \sigma} + h.c. ).
\end{equation}
 The first term is the hybridization term, which is equivalent to that of the Kondo model.
 We can identify that $\rho_K = V X^0_i/J_K = \frac{U}{4 V} X^0_i$. The second term now
 provides the incoherent fluctuations, which, as we argue in the main text, can mediate the same
 chiral interactions for the $d$-electrons through the triangular diagrams. But in this approach,
 the $X$-field satisfies a spectral sum rule: $\int d\nu d^2k / (2 \pi)^3 G_X(\nu;{\bf k}) = 1$,
 from which we can obtain that in the Kondo screened phase
 \begin{equation}
   H_{\text{chiral}} = \left(1 - \frac{4J_K}{U} \rho_K ^2 \right)^3 \frac{J_K^3}{2} E_{ijk} \s_i \cdot (\s_j \times \s_k).\label{eq:S-12}
 \end{equation}
Note that the prefactor $4J_K/U $ is changing as we tune $J_K$. In our calculation, we fix $U = 16t$,
i.e. twice as the $d$-electron's bandwidth.

{\it Hubbard-Stratonovich transformation of $H_{\text{chiral}}$}.
In this part, we construct a Hubbard-Stratonovich (HS) transformation to decouple the six-fermion chiral interaction of Eq. (\ref{eq:S-12}). With such a HS transformation, we apply a gauge fixing condition so that the time-reversal-symmetry breaking phases are carried by the fermion bilinears, which yields Eq. (8) of the main text.

In general, we need to introduce two sets of Hubbard-Stratonovich (HS) fields, namely, $\gamma$s, $\kappa$s, which can be interpreted as a single bond / two consecutive bonds fields:
\begin{eqnarray}
  \label{eq:HS-fields}
  \gamma_{ij} & = \ev{\sum_{\alpha} d^\dagger_{i,\alpha} d_{j,\alpha} }, \\
  \kappa_{ij,k} & = \ev{\sum_{\alpha,\beta} d^\dagger_{i,\alpha} d_{k,\alpha} d^\dagger_{k,\beta} d_{j,\beta} },
\end{eqnarray}
which are, in principle, independent.
The bond indices here are directional, i.e. $\gamma_{ji} = \gamma_{ij}^* $, $\kappa_{ji,k} = \kappa_{ij,k}^*$.
There are 6 complex $\gamma$'s and 12 complex $\kappa$'s.

The HS transformation is as follows
\begin{equation}
  \label{eq:HS-action}
  L_{\text{HS}} = \sum_{\x} (d^\dagger_{\x}(i\pt_t + \mu)d_{\x} - \widetilde{\Psi}^* \mathcal{M} \Psi  + J^\dagger \Psi + \widetilde{\Psi}^* J - H_{\text{kin}} ),
\end{equation}
where $\widetilde{\Psi}^{*} = \{\kappa^*_{\x_i \x_j,k},\dots \kappa^*_{\x_i \x_j,\bar{k}},\dots, \gamma^*_{\x_i \x_j},\dots,\gamma_{\x_j \x_i}^*,\dots \}$ is a 24-component vector. The indices $i-j$ run over all the links inside a unit cell given by $\x$, and $\{k, \bar{k}\}$ denote the other two sites for a given bond $\ev{ij}$ within the unit cell.
\begin{equation}
  \begin{split}
  &   J^\dagger = \{ \hat{\gamma}^\dagger_{ij}, \hat{\gamma_{ji}}^\dagger, \dots, {\kappa}_{ij,k}^\dagger, \hat{\kappa}_{ij,\bar{k}}^\dagger, \dots \}.
  \end{split}
\end{equation}
where we use $\hat{\gamma}_{ij}^\dagger = \sum_{\alpha} d^\dagger_{\x_j,\alpha} d_{\x_i,\alpha}$, $\hat{\kappa}^\dagger_{ij,k} = \sum_{\alpha,\beta} d^\dagger_{j,\alpha} d_{k,\alpha} d^\dagger_{k,\beta} d_{i,\beta}$.

To determine $\mathcal{M}$, suppose that we now integrate out all the HS fields, we should recover the effective interactions as
\begin{equation}
 H_{eff-int} = J^\dagger \mathcal{M}^{-1} J = H_{\text{cc}} + \dots,
\end{equation}
in which the $\dots$ indicates other effective interactions. To have a stable HS transformation, we need to include further the 4-fermion effective interactions at $H^{(2)}_{eff-int} \sim \mathcal{O}(J_K^2)$ generated from $J_K^2  (\s_i\cdot \S_i) (\s_j\cdot \S_j)$ as well as the 8-fermion process at $H^{(4)}_{eff-int}\sim \mathcal{O}(J_K^4)$. Since the $f$-fermions are gapped, we can keep only the short-range terms, i.e. within a unit cell, so that all the terms can be decoupled by the $\hat{\gamma}_{ij}$s and $\hat{\kappa}_{ij}$s in the large-$N$ limit. Then $\mathcal{M}^{-1}$ can be written in a block form $\mathcal{M}^{-1} =  \oplus (\mathcal{M}^{-1}_{(ij)})$, where $(\mathcal{M}^{-1}_{(ij)})$ is a $4\times 4$ matrix for a given bond $(ij)$ within the unit cell.

Here we estimate the matrix elements of $(\mathcal{M}^{-1}_{(ij)})$ within the approximations that are
used for computing $H_{\text{chiral}}$, i.e., equal-time contraction is used and only those within a unit cell are included:
\begin{eqnarray}
  \label{eq:M-inv}
  & (M^{-1})_{\hat{\gamma}^{\dagger} \hat{\gamma},(ij)} = J_K^2 \text{sgn}[(ij)] \chi_{ij}\chi_{ji} /2! = \rho_{ij}^2 ,\\
  & (M^{-1})_{\hat{\kappa} \hat{\gamma},(ij)k} = J_K^3 P_{ijk} / (2\times 3!), \\
  &  (M^{-1})_{\hat{\kappa}^\dagger \hat{\kappa},(ij),k k'} = \delta_{k,k'} J_K^4 \rho_{ij}^4 / 4!.
\end{eqnarray}
$\text{sgn}[\ev{ij}] $ is a relative sign coming from the fact that $\hat{\gamma_{ij}}^\dagger \hat{\gamma_{ij}}
\sim - \hat{\gamma_{ji}}^\dagger \hat{\gamma_{ji}}$.
We see that $\det [M^{-1}]$ is indeed positive; hence, this is a stable HS transformation.
$\mathcal{M}$ is then obtained by inverting $\mathcal{M}^{-1}$.

Therefore, we have a formal HS decoupling of $H_{\text{chiral}}$. Further replacing the HS-fields
by their expectation values in Eq. (\ref{eq:HS-action}), we obtain both fermion bilinears and four-fermion terms.
To lower the total energy, we need to have $\gamma_{ij} \kappa^*_{ij,k} \sim - P_{ijk}$. Upon satisfying this constraint,
we have an additional gauge degree of freedom to choose either $\gamma_{d,ij}$ or $\kappa^*_{ij,k}$ to be imaginary,
i.e. explicitly breaking TRS, even though the underlying physical state is the same.
For convenience, we can choose $\kappa^*_{ij,k}$s which couple to $d$-fermion bilinears ($\hat{\gamma}_{ij}$s) to be TRSB.
By keeping only the TRSB terms in Eq. (\ref{eq:HS-action}), we justify our choice of Eq. (8) in the main text as
\begin{equation}
  \label{eq:Hd1}
  H_{d,1} = \sum_{\ev{ij},k} (\kappa^*_{ij,k} \hat{\gamma}_{ij} + h.c.).
\end{equation}

\subsection{Berry curvature, Berry connection, Streda formula and Kubo formula}

The AHE coefficient, $\sigma_{xy}$, presented in the main text are computed using the Streda formula:
\begin{equation}\label{eq:streda-formula}
  \begin{split}
    \sigma_{xy}  &= \int \frac{d{\bf k}}{(2\pi)^2} \sum_{n} \mathcal{F}^{xy}_n({\bf k}) f(\epsilon_n({\bf k}))\\
    & =  \sum_{n \neq n'} \int \frac{d{\bf k}}{(2\pi)^2}
    [f(\epsilon_n({\bf k})) - f(\epsilon_{n'}({\bf k}))] \\
    & \times \text{Im} \frac{\meanvalue{n,{\bf k}}{v_x({\bf k})}{n',{\bf k}}  \meanvalue{n',{\bf k}}{v_y({\bf k})}{n,{\bf k}} }
    {[\epsilon_n({\bf k}) - \epsilon_{n'}({\bf k})]^2} \quad .
  \end{split}
\end{equation}
Here, $v_a({\bf k}) = \pt_a H_{d}({\bf k})$ is
the current operator of the conduction electrons,
$\mathcal{F}^{xy}_n({\bf k}) $ the Berry curvature, and $f(\epsilon_n({\bf k}))$ the Fermi function.
Both $\hbar$ and $e$ have been taken to be $1$.

To discuss the role of the Berry curvature, we start from the more standard Kubo formula. The current operators are
\begin{equation}
  \bm{J}_{\q} = \frac{1}{\sqrt{N}} \sum_{{\bf k}} c^\dagger_{{\bf k}+\q/2} \frac{\pt H_k}{\pt {\bf k}} c_{{\bf k}-\q/2}.
\end{equation}
In frequency-momentum space, the conductivity is computed via the current-current correlation function
\begin{equation}\label{eq:sigma-xy}
  \begin{split}
  &  \pi_{ab}(i\nu)= \sum_{\omega} \int \frac{d^2k}{(2\pi)^2}\tr\Big[\frac{\pt H}{\pt k_a}\\
& \times G(\omega -\nu,{\bf k}-\q/2)\frac{\pt
      H}{\pt k_b}G(\omega,{\bf k} + q/2) \Big],
  \end{split}
\end{equation}
where the sum over $\omega$ is Matsubara sum.
\begin{equation}
\sigma_{ab} =  \lim_{\omega \rightarrow 0} \Big[- \im [\pi_{ab}(i\nu)/\nu\big\vert_{i \nu\rightarrow \omega + i 0^+}] \Big].
\end{equation}
For convenience, it is better to write both $G$ and $H$ in terms of the Bloch bands projection
operators $P_n({\bf k}) = \proj{n, {\bf k}}{n, {\bf k}}$ (which is possible for fermion bilinear theory) with $\ket{n,{\bf k}}$
being the eigenvectors of $n$th band at momentum ${\bf k}$:
\begin{eqnarray}
  H({\bf k}) & = \sum_n \epsilon_n({\bf k}) P_n({\bf k}), \\
  G(\omega,{\bf k}) & = \sum_n \frac{P_n({\bf k})}{i\omega - \epsilon_n({\bf k}) }.
\end{eqnarray}
After inserting the expression into Eq. (\ref{eq:sigma-xy}), we find that only the following term contributes
\begin{equation}
  \begin{split}
    &\pi_{ab}(i\nu) = \sum_{\omega} \int \frac{d^2k }{(2\pi)^2} \sum_{n_0,\dots,n_3} \tr \Big[\pt_{k_a} P_{n_0}\\
    & \times P_{n_1} \pt_{k_b} P_{n_2} P_{n_3} \frac{\epsilon_{n_0} \epsilon_{n_2}}{(i(\omega-\nu)
    -\epsilon_{n_1}) (i\omega - \epsilon_{n_3})}\Big]\\
    & =  \int \frac{d^2k }{(2\pi)^2} \sum_{n_0,\dots,n_3} \tr \Big[\pt_{k_a} P_{n_0}\\
& \times P_{n_1} \pt_{k_b} P_{n_2} P_{n_3} \frac{\epsilon_{n_0} \epsilon_{n_2}(f(\epsilon_{n_3})
- f(\epsilon_{n_1}))}{i\nu + \epsilon_{n_3} - \epsilon_{n_1}}\Big].
  \end{split}
\end{equation}
Here $f(\epsilon)$ is the Fermi distribution function and arises from the Matsubara sum. The sum of $n_i$
runs over band indices. After performing the $\tr$ operation, we end up with the following result
\begin{equation}
   \begin{split}
    &\pi_{ab}(i\nu) = \int \frac{d^2k }{(2\pi)^2} \sum_{n,n'}\Big[- A_{nn'}^a A_{n' n}^b  \frac{f(\epsilon_n)
    - f(\epsilon_{n'})}{i\nu + \epsilon_{n} - \epsilon_{n'}}  \epsilon_{n} \epsilon_{n'}\\
& + A^a_{nn'} A^{b*}_{n'n} \frac{f(\epsilon_n) - f(\epsilon_{n'})}{i\nu + \epsilon_{n} - \epsilon_{n'}} \epsilon^2_{n'}
+ A^{a*}_{nn'} A^{b}_{n'n} \frac{f(\epsilon_n) - f(\epsilon_{n'})}{i\nu + \epsilon_{n} - \epsilon_{n'}} \epsilon^2_{n}\\
& -  A_{nn'}^{a*} A_{n' n}^{b*}  \frac{f(\epsilon_n) - f(\epsilon_{n'})}{i\nu + \epsilon_{n} - \epsilon_{n'}}
\epsilon_{n} \epsilon_{n'}\Big],
\end{split}
\end{equation}
where $A_{n,n'}^a = -i \braket{n}{\pt_{k_a} n'}$, $A_{n,n'}^{a*} = -i \braket{\pt_{k_a} n}{n'}$ is the matrix
element of $\pt_{k_a}$. Note only the diagonal elements are the Berry connection.
Then we perform an analytic continuation $i\nu \rightarrow \omega + i \eta$, and take the imaginary part of
$\pi_{ab}(\omega)/\omega$. In the end, we let $\omega \rightarrow 0$. When we take the imaginary part
of $\pi_{ab}(\omega)/\omega$, we have two different contributions:
\begin{equation}
   \begin{split}
    & \pi_{ab}^{(1)} = \int  \frac{d^2k }{(2\pi)^2} \sum_{n,n'} \Big( \im [- A_{nn'}^a A_{n' n}^b] \\
    &\times \re [\frac{f(\epsilon_n) - f(\epsilon_{n'})}{\omega + \epsilon_{n} - \epsilon_{n'} + i \eta}
     \epsilon_{n} \epsilon_{n'}] + \dots\Big),
  \end{split}
\end{equation}
\begin{equation}
\begin{split}
  & \pi_{ab}^{(2)} =  \int  \frac{d^2k }{(2\pi)^2} \sum_{n,n'}  \Big(  \re [- A_{nn'}^a A_{n' n}^b] \\
& \times \im[\frac{f(\epsilon_n) - f(\epsilon_{n'})}{\omega + \epsilon_{n} -
    \epsilon_{n'} + i \eta} \epsilon_{n} \epsilon_{n'}] + \dots \Big),
\end{split}
\end{equation}
where $\dots$ denotes the rest three terms. Note that $A_{nn'}^a = - A_{nn'}^{a*}$, $(A_{nn'}^a)^*
= A_{n'n}^{a*}$, we find
\begin{equation}
  \begin{split}
    & \pi_{ab}^{(1)} = \int  \frac{d^2k }{(2\pi)^2} \sum_{n,n'} \Big( \im [- A_{nn'}^a A_{n' n}^b] (f(\epsilon_n)
    - f(\epsilon_{n'}))\\
    & \times \re [\frac{(\epsilon_{n} - \epsilon_{n'})^2}{\omega + \epsilon_{n} - \epsilon_{n'} + i \eta}
    + \frac{(\epsilon_{n} - \epsilon_{n'})^2}{\omega - \epsilon_{n} + \epsilon_{n'} + i \eta}] \\
& = \int  \frac{d^2k }{(2\pi)^2} \sum_{n,n'} \Big( \im [- A_{nn'}^a A_{n' n}^b] (f(\epsilon_n) - f(\epsilon_{n'})) \\
& \times  \frac{2 \omega (\epsilon_{n} - \epsilon_{n'})^2}{\omega^2 - (\epsilon_{n} - \epsilon_{n'})^2}.
  \end{split}
\end{equation}
Therefore, after taking the limit $ \pi_{ab}^{(1)} / \omega|_{\omega \rightarrow 0}$, we obtain
\begin{equation}\label{eq:sigma1}
  \sigma^{(1)}_{ab} = \int  \frac{d^2k }{(2\pi)^2} \sum_{n \neq n'} \im [A_{nn'}^a A_{n' n}^b] (f(\epsilon_n)
  - f(\epsilon_{n'})),
\end{equation}
which we can use the relation $\meanvalue{n}{\pt_{k_a} H(k)}{n'} = (\epsilon_{n} - \epsilon_{n'}) A_{nn'}^a
+ \delta_{n,n'} \pt_{k_a} \epsilon_n$ to transform into the Streda formula.
For $\pi_{ab}^{(2)}/\omega|_{\omega \rightarrow 0}$,
\begin{equation}
\im[\frac{f(\epsilon_n) - f(\epsilon_{n'})}{\omega + \epsilon_{n} - \epsilon_{n'} + i \eta}] = \delta_{n,n'}
 \delta(\omega) \frac{f(\epsilon_n) - f(\epsilon_{n} + \omega)}{\omega}|_{\omega \rightarrow 0}.
\end{equation}
Note that the factor is immediately zero for $n \neq n'$ when we take $\eta \rightarrow 0$. Only $n = n'$ terms survive, and $\pi_{ab}^{(2)}/\omega|_{\omega \rightarrow 0} \sim  \delta(\omega) \pt f(\epsilon)/\pt \epsilon$. For $a\neq b$, $ \re[A_{nn}^a A_{n n}^b]$ is symmetric upon exchanging $a \leftrightarrow b$. On the other hand, we know $\pi_{ab} = - \pi_{ba}$. Therefore, $\pi_{ab}^{(2)} = 0$ for $a \neq b$. When $a = b$, we recover the usual Kubo formula result of dc conductivity $\sigma_{xx} \sim \delta(\omega)$.

To show that $\sigma_{xy}^{AH}$ is a Fermi surface property\cite{Haldane2004S}, we can rewrite Eq. (\ref{eq:sigma1})
through an integration
by part, and make use the fact \cite{Sun2008} that
$\sum_{n'}\im [A_{nn'}^a A_{n' n}^b]
= \nabla_{{\bf k}}^a A_{nn}^b - \nabla_{{\bf k}}^b A_{nn}^a$:
\begin{equation}
  \begin{split}
  & \sigma^{(1)}_{ab} = \int \frac{d^2k }{2\pi^2}  \sum_n
    f(\epsilon_n) (\nabla_{{\bf k}}^a A_{nn}^b - \nabla_{{\bf k}}^b A_{nn}^a) \\
    & = \int \frac{d^2k }{2\pi^2}  \sum_n (A_{nn}^b \nabla_{{\bf k}}^af(\epsilon_n) - A_{nn}^a \nabla_{{\bf k}}^bf(\epsilon_n) ) \\
& = \sum_n \frac{1}{2 \pi^2}  \oint A^a_n({\bf k}_F) d{\bf k}_{F a}.
  \end{split}
\end{equation}

\subsection{Berry curvature distribution}
\begin{figure}
  \centering
   \subfigure[]{\label{fig:BerryCurvature1}\includegraphics[width=4cm]{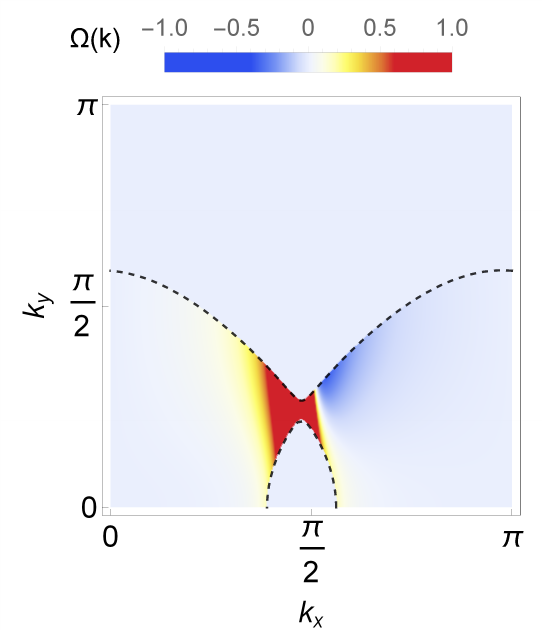}}
   \subfigure[]{\label{fig:BerryCurvature2}\includegraphics[width=4cm]{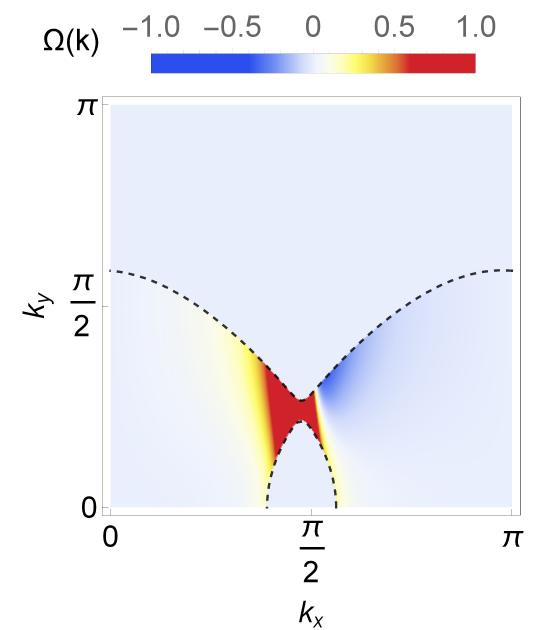}}
   \subfigure[]{\label{sfig:BCFS_KD}\includegraphics[width=0.45\columnwidth]{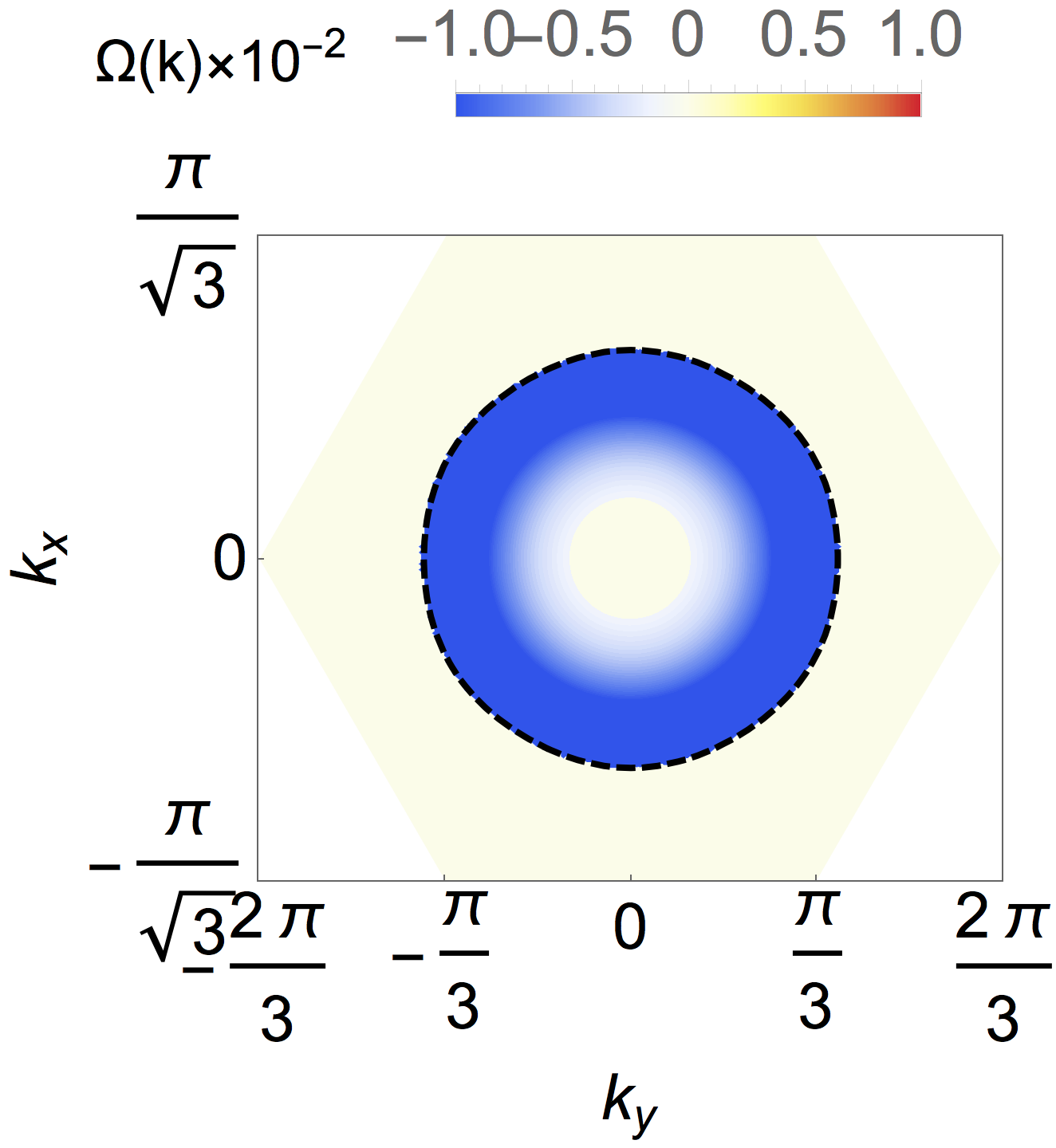}}~~~
   \subfigure[]{\label{sfig:BCFS_KS}\includegraphics[width=0.45\columnwidth]{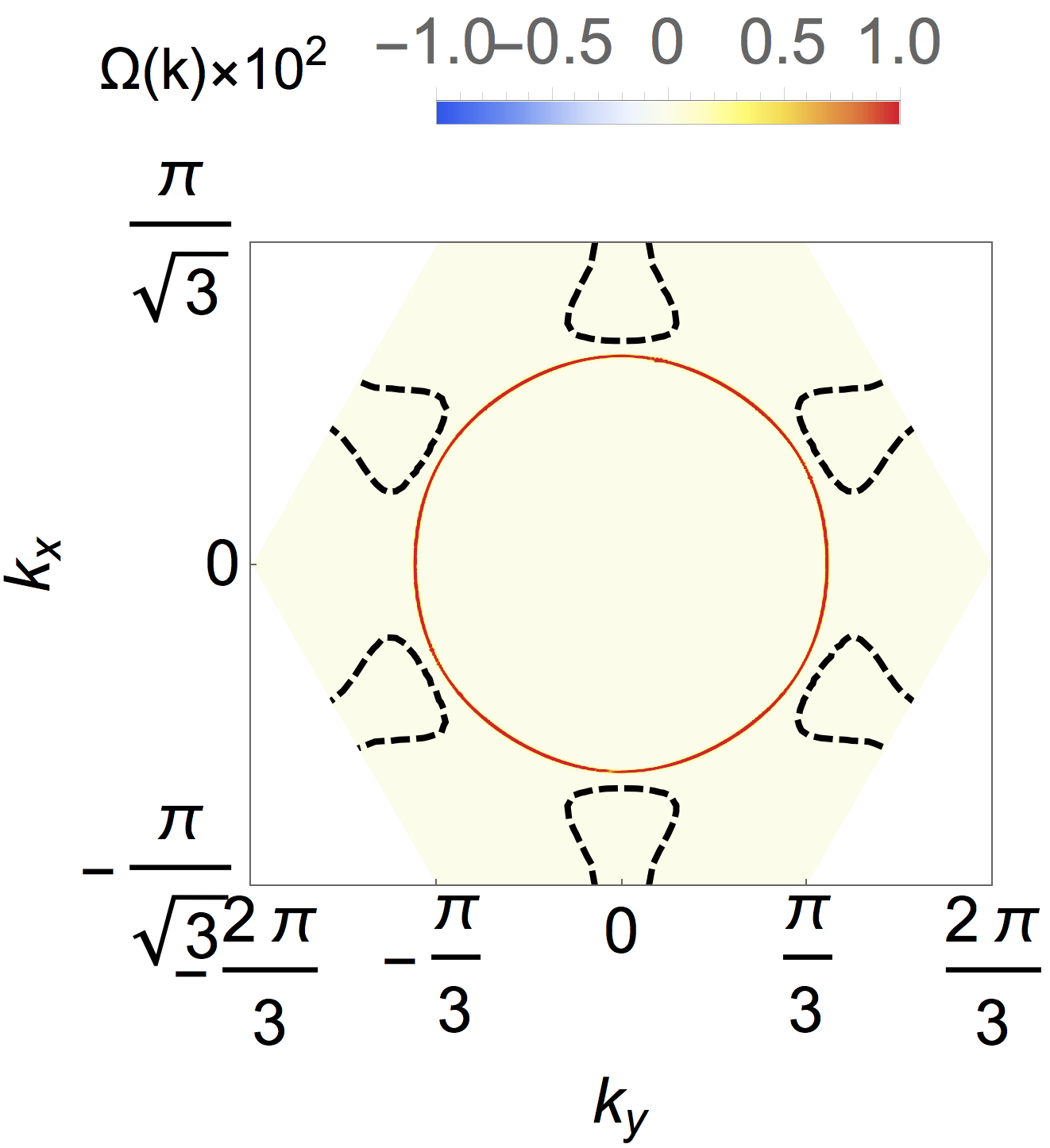}}\\
   \caption{(Color online)
   The Berry curvature distributions for the square lattice model, $P_S$ phase (\ref{fig:BerryCurvature1}) and $P_L$ phase (\ref{fig:BerryCurvature2});
   and for the kagom\'{e} lattice model,  $P_S$ phase (\ref{sfig:BCFS_KD}) and $P_L$ phase (\ref{sfig:BCFS_KS}). The Fermi surfaces are also shown as black dashed lines.}\label{sup-fig:BC-dis}
 \end{figure}

 The Berry curvature distributions are shown in Fig.~(\ref{sup-fig:BC-dis}) for the $P_S$ phase (S\ref{fig:BerryCurvature1}) and $P_L$ phase (S\ref{fig:BerryCurvature2}) of the square lattice model, as well as the $P_S$ phase (S\ref{sfig:BCFS_KD}) and $P_L$ phase (S\ref{sfig:BCFS_KS}) of the kagom\'{e} lattice model. For (S\ref{fig:BerryCurvature1}) and (S\ref{fig:BerryCurvature2}), despite the visual resemblance, their difference is still significant as shown in
 Fig.~(3a) of the main text.

 \subsection{Reconstruction of Fermi Surfaces}

We note that in the Kondo-destroying $P_S$ phase, only the conduction electrons participate in forming the Fermi surface. By contrast,
in the Kondo-screened $P_L$ phase,
both the conduction electrons and local $f$-moments are involved in forming the Fermi surface~\cite{Oshikawa2000}.
In the case of the $J_1-J_2$ model, the spinon excitations of the CSL phase are gapped. By contrast, in the kagom\'{e} case, they are gapless.

\begin{figure*}
  \centering
 \subfigure[]{\label{sfig:j1j2DOS}\includegraphics[width=7.8cm]{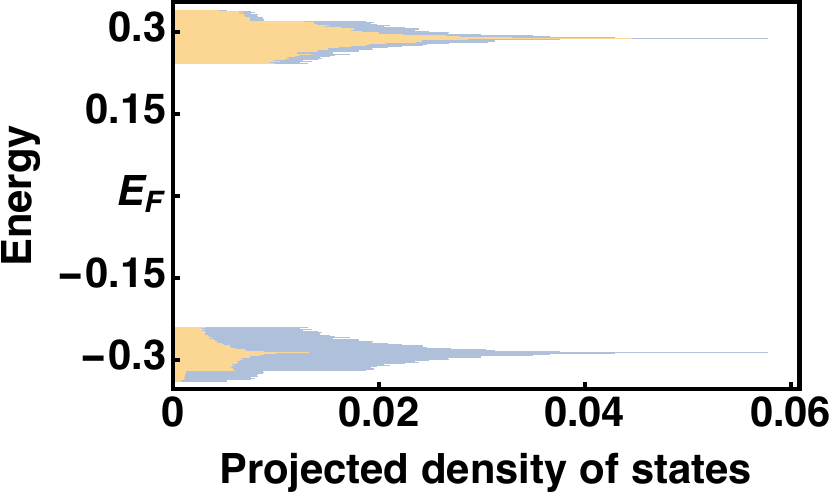}}
  \subfigure[]{\label{sfig:j1j2DOS_leg}\includegraphics[width=.6cm]{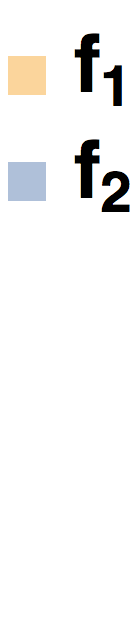}}
 \subfigure[]{\label{sfig:kagDOS}\includegraphics[width=7.4cm]{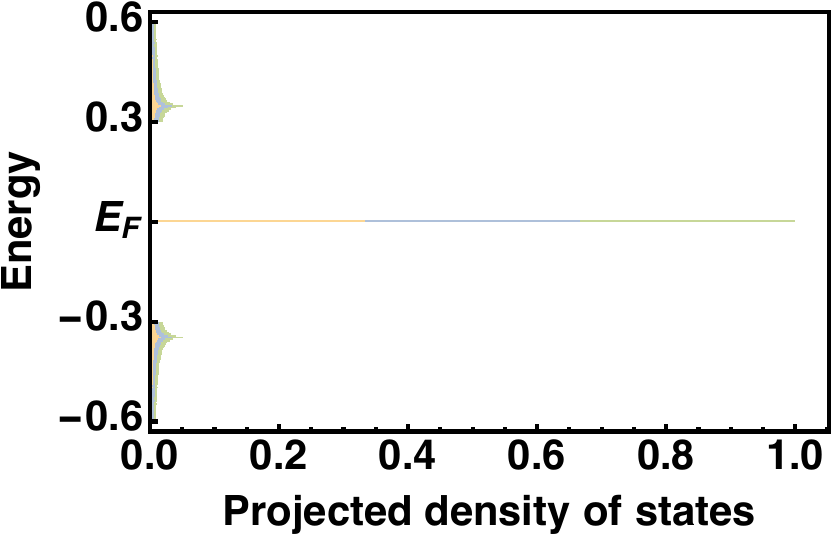}}
   \subfigure[]{\label{sfig:kagDOS_leg}\includegraphics[width=.6cm]{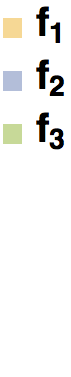}}
  \subfigure[]{\label{sfig:j1j2DOS_KS}\includegraphics[width=7.4cm]{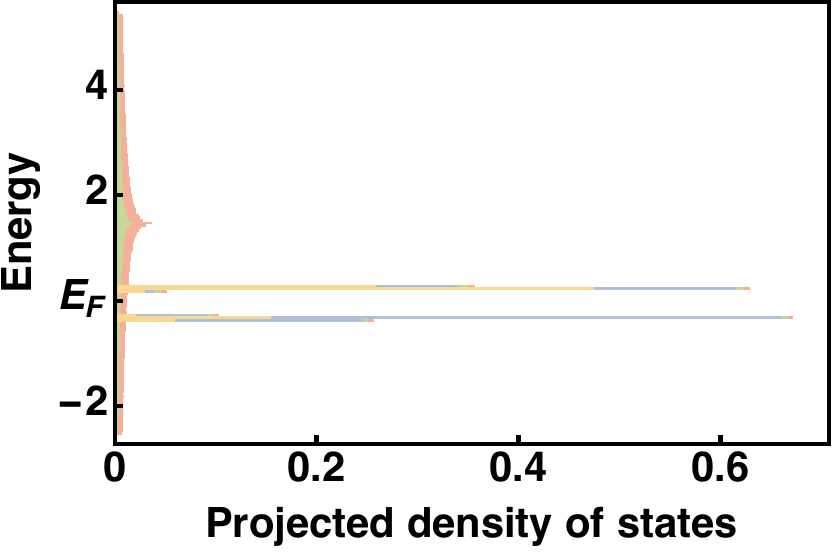}}
   \subfigure[]{\label{sfig:j1j2DOS_KS_leg}\includegraphics[width=.6cm]{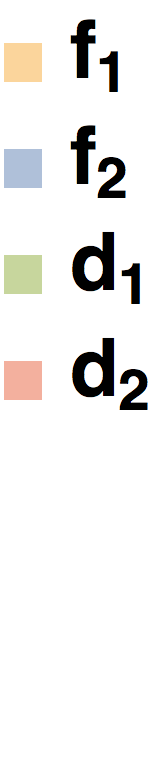}}
 \subfigure[]{\label{sfig:kagomeDOS_KS}\includegraphics[width=7.4cm]{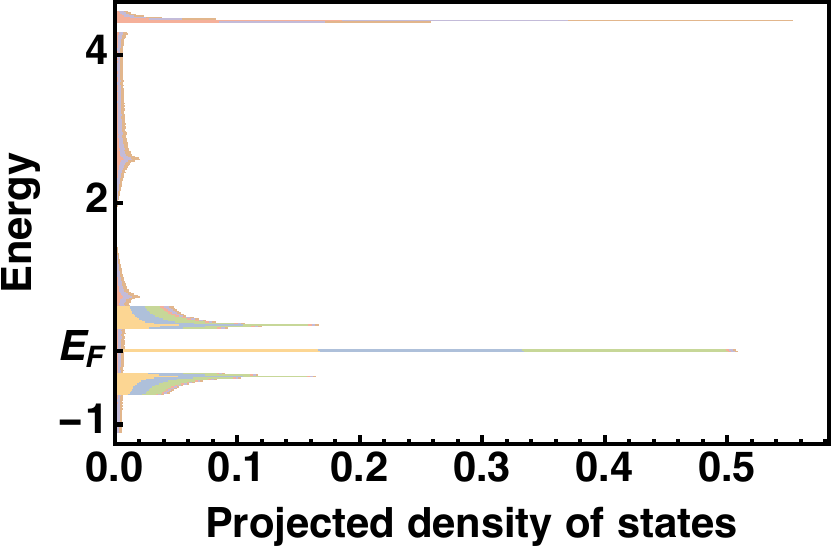}}
   \subfigure[]{\label{sfig:kagomeDOS_KS_leg}\includegraphics[width=.6cm]{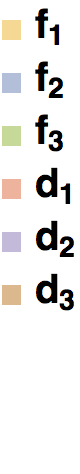}}
  \caption{Density of states projected to the sites of a unit cell, for
  (a) $J_1-J_2$ model's spinons in its CSL state,
  (c) kagom\'{e} lattice model's spinons in the ($\frac{\pi}2,-\pi$) state,
  (e) Kondo screened phase of the $J_1-J_2$ model,
  (g) Kondo screened phase of the kagom\'{e} lattice model;
  (b),(d),(f), and (h) shows the relevant legends for color corresponding to the original eigenfunction elements.} \label{fig:DOS}
\end{figure*}

To illustrate the point,
we show the projected density of states (DOS) in Fig.~(\ref{fig:DOS}).
The parent spinon TRSB flux state is gapped at
zero energy  (referred to as ``$E_F$")
for the $J_1-J_2$ case, but is
 gapless at $E_F$ for the
Kagom\'{e} case.
The DOS structure of the spinons survives the $P_L$ phase (bottom row),
but are constrained to straddle $E_F$.
The Fermi surface is only affected in the Kagom\'{e} lattice.

\begin{figure}
  \centering
  \subfigure[]{\label{sfig:sq-kd-fs}\includegraphics[width=.45\columnwidth]{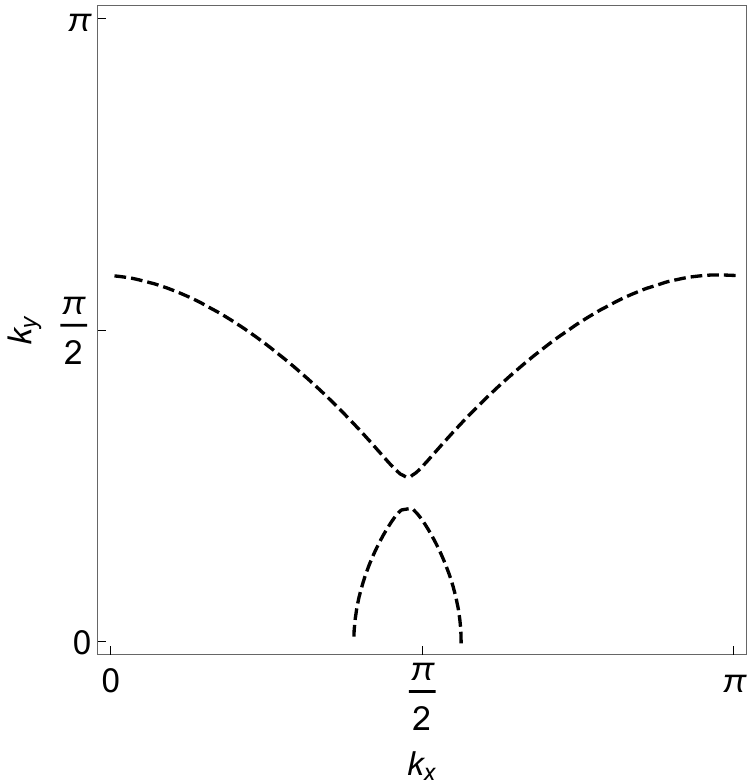}}
  \subfigure[]{\label{sfig:sq-ks-fs}\includegraphics[width=.45\columnwidth]{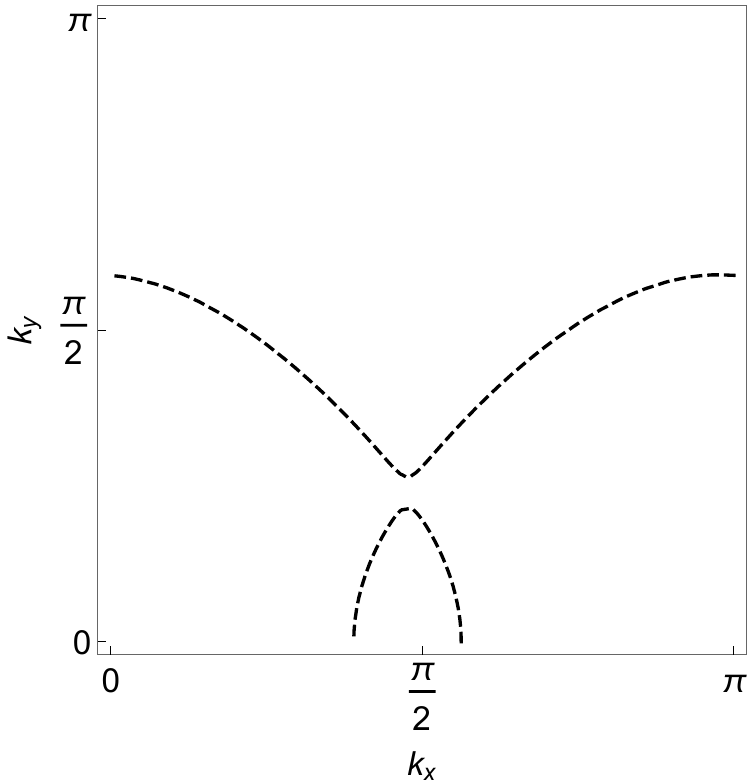}}
  \subfigure[]{\label{sfig:kg-kd-fs}\includegraphics[width=.45\columnwidth]{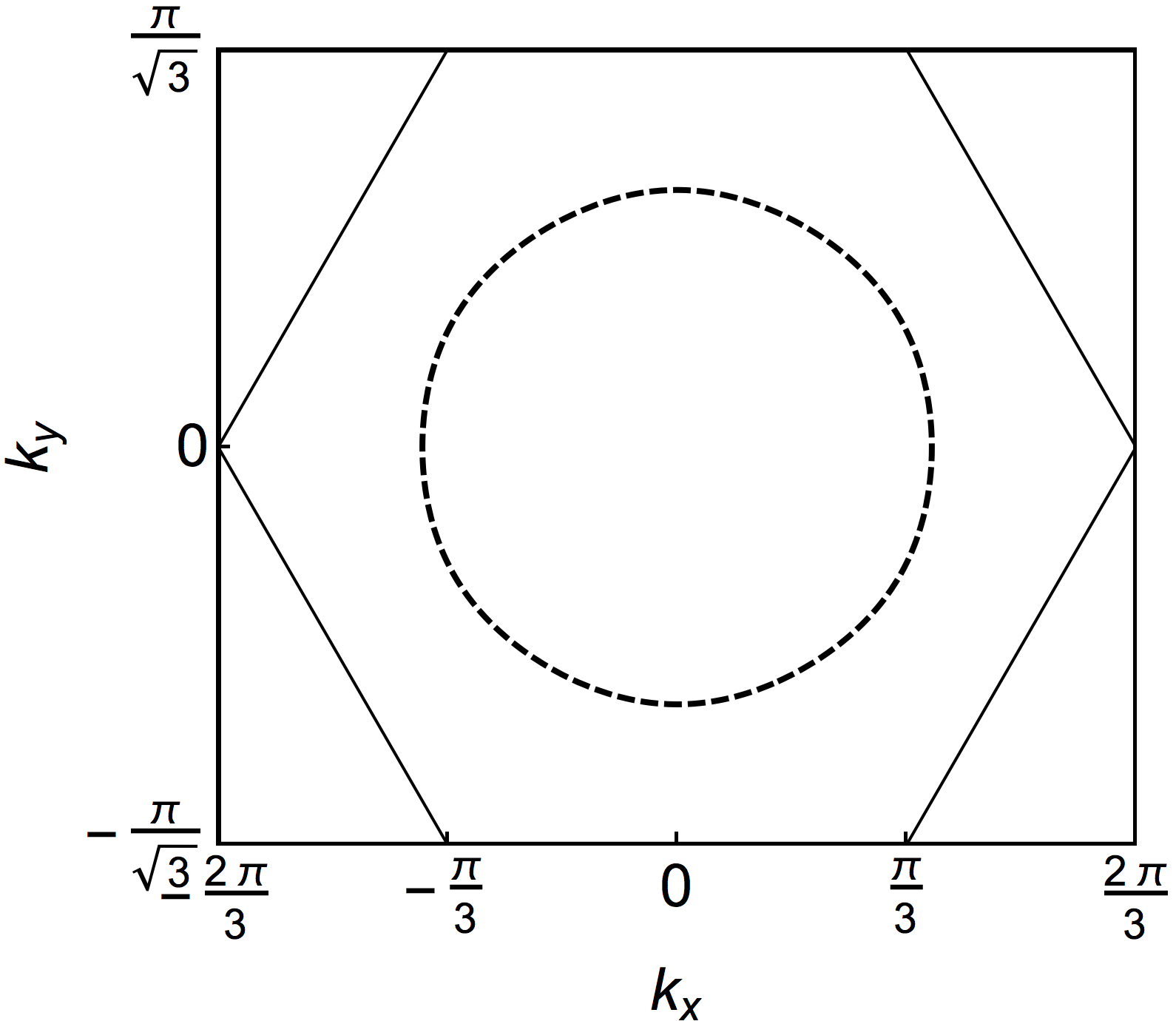}}
  \subfigure[]{\label{sfig:kg-ks-fs}\includegraphics[width=.45\columnwidth]{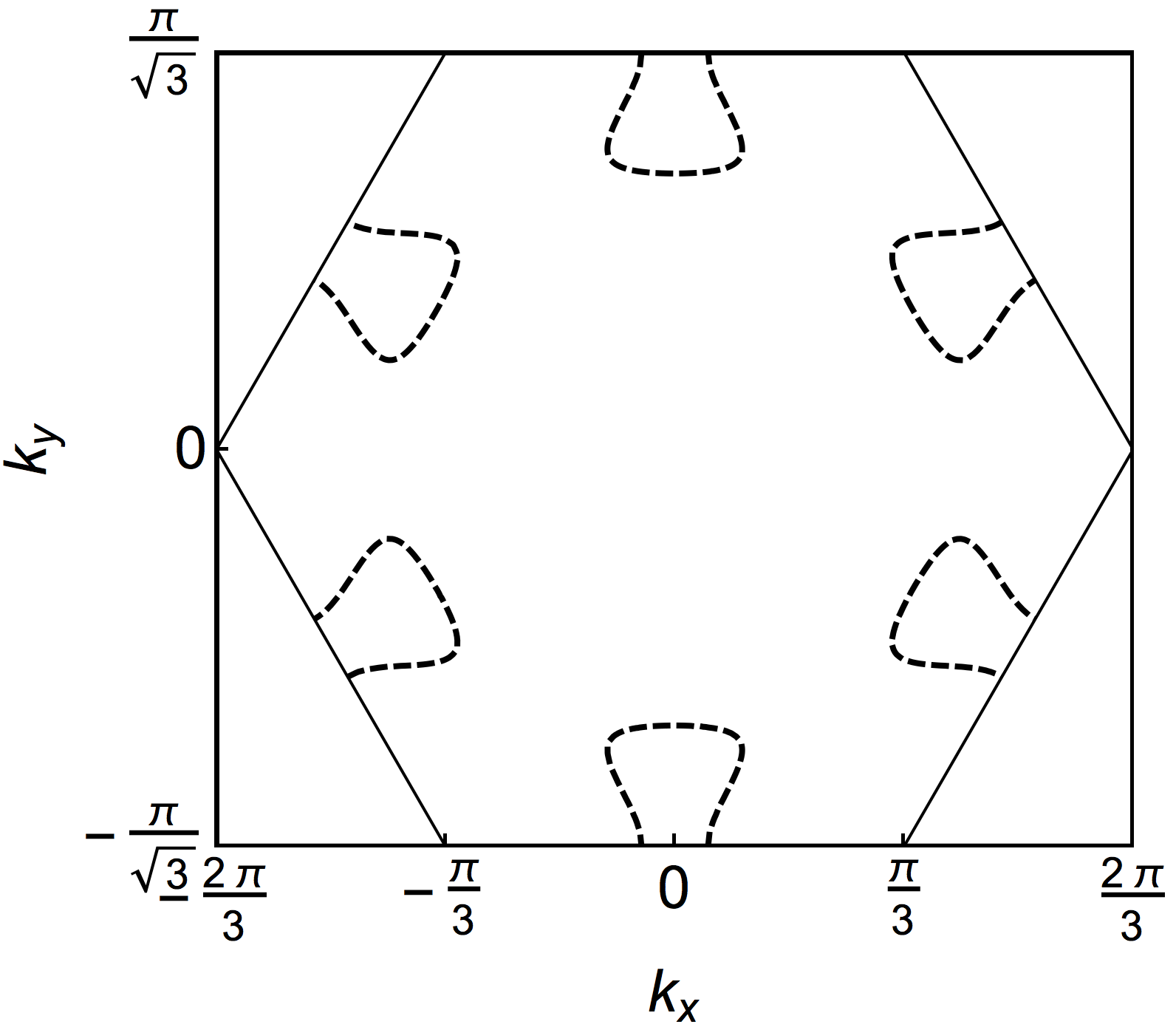}}
  \caption{The Fermi surfaces of the square lattice model in the $P_S$ phase (\ref{sfig:sq-kd-fs}) and in the $P_L$ (\ref{sfig:sq-ks-fs}) phase, and of the Kagom\'{e} lattice model in the $P_S$ phase (\ref{sfig:kg-kd-fs}) and in the $P_L$ phase (\ref{sfig:kg-ks-fs}).}\label{fig:fs}
\end{figure}
This can be seen by directly plotting the Fermi surfaces. Fig.~(\ref{fig:fs}) shows
both the Kondo-destroyed and the Kondo-screened phases
for both the square lattice model and the kagom\'{e} lattice model. It is seen that, for the $J_1-J_2$ model on the square lattice,
the Fermi surface smoothly evolves through the QCP. By contrast, for the kagom\'{e} lattice,
the Fermi surface experiences a sudden jump across the QCP. We also note that the jump is very substantial. This is because, in
the kagom\'{e} lattice's CSL state, the middle spinon band
happens to be a flat band.

\subsection{Analysis of the wavefunction reconstruction across the QCP}
To further our understanding
about the nonanalyticities across the QCP,
we rewrite the Hamiltonian across the QCP in terms of the $d$-band and $f$-band eigenstates,
which we denote as $\ket{\phi^d_{\k}}$ and $\ket{\phi^f_{\k}}$ respectively:
\begin{equation}
  \label{eq:H-phi-basis}
  \begin{split}
    & H = \\
    & \sum_{\k}
  \begin{pmatrix}
    (\epsilon^d_{\k}-\mu + \lambda')  \ket{\phi^d_{\k}} \bra{\phi^d_{\k}} & \delta \ket{\phi^d_{\k}} \bra{\phi^f_{\k}}  \\
    \delta \ket{\phi^f_{\k}} \bra{\phi^d_{\k}} & (\epsilon^f_{\k}-\mu - \lambda')  \ket{\phi^f_{\k}} \bra{\phi^f_{\k}}
  \end{pmatrix}.
  \end{split}
\end{equation}
Here, $\delta$ is the hybridization strength.
In addition,
$\lambda'$ is the Lagrangian multiplier,
which is shifted from $\lambda$
by a constant that can be absorbed into $\mu$, to obtain the above symmetric form for later convenience.
The hybridization, thus the wavefunction reconstruction,
is the strongest at the $\k$ points where the conduction bands and spinon bands intersect, i.e., $\epsilon^f_{\k_0} -\mu = \epsilon^d_{\k_0}  -\mu = 0$.

For the kagom{\'e} case,
consider the case that the Fermi surface jumps.
We expect $\lambda'$
to track  $\delta$ as the QCP is approached.
Nonetheless, we can still start out with the points where $\epsilon^f_{\k_0} -\mu = \epsilon^d_{\k_0}  -\mu = 0$. In this case,
we can write the $\lambda'$ term as $\lambda' \sigma_0\otimes \tau_z$, where $\tau_z$ is the Pauli matrix for the orbital space.
This term does not commute with the hybridization term, which is off-diagonal in the $\tau$ space.
(Note that
both  the diagonal and
 off-diagonal blocks above
are diagonal matrices in the sublattice space, and therefore commute with each other in that space.)
Therefore, the presence of any $\lambda'$ prevents us from block-diagonalizing the Hamiltonian even for infinitesimal
$\delta$.
The new eigenstates are, therefore, reconstructed completely.

\begin{figure}[b!]
  \centering
   \includegraphics[width=0.75\columnwidth]{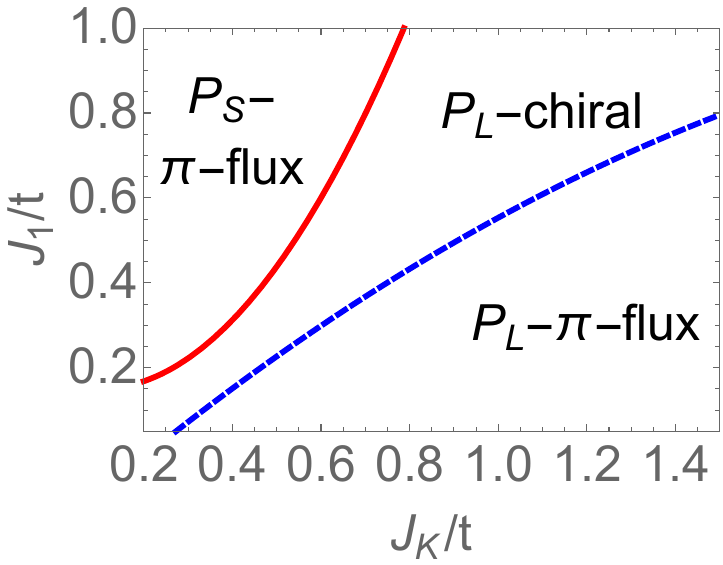}\\
  \caption{(Color online)
The phase diagram of the $J_1 - J_2 -J_K$ model with $J_2=J_1/2$ and $n_d = 0.5$ (\ref{fig:large-N-pd}).
}
\label{fig:large-N-pd}
\end{figure}

\subsection{Phase diagram in the saddle-point analysis}
\label{sec:phase-diagram-saddle}
To illustrate our procedure, we consider the phase diagram arising from the saddle-point analysis in the case of $J_1-J_2$ square lattice.
We minimize the total energy of Eq.~(2)
 with respect to the amplitudes of the link fields $\rho_{ij}$ and $\rho_{K,i}$.
The phase diagram of the square lattice model is shown in Fig.~(\ref{fig:large-N-pd}),
where the red (solid) and blue (dashed) lines, respectively, mark a first-order
phase transition and a crossover.
It shows that both the flux-state and the chiral-state solutions can be stabilized,
i.e.\ having lower energies than the unhybridized phase,  for $J_K$ larger than some critical $J_{K,c}$.
The flux phase solution has the lowest energy when stabilized, signaling that the Kondo coupling favors
the gapless states.

For the pyrochlore lattice, the CSL state is gapless~\cite{Burnell2009s}, and our result here strongly suggests that a similar chiral state
could be the ground state on the pyrochlore lattice when the Kondo coupling is introduced.

\subsection{Potential experimental realization of the frustrated $J_1 - J_2$ Kondo lattice}

\begin{figure}[t!]
\centerline{\includegraphics[width=0.4\columnwidth]{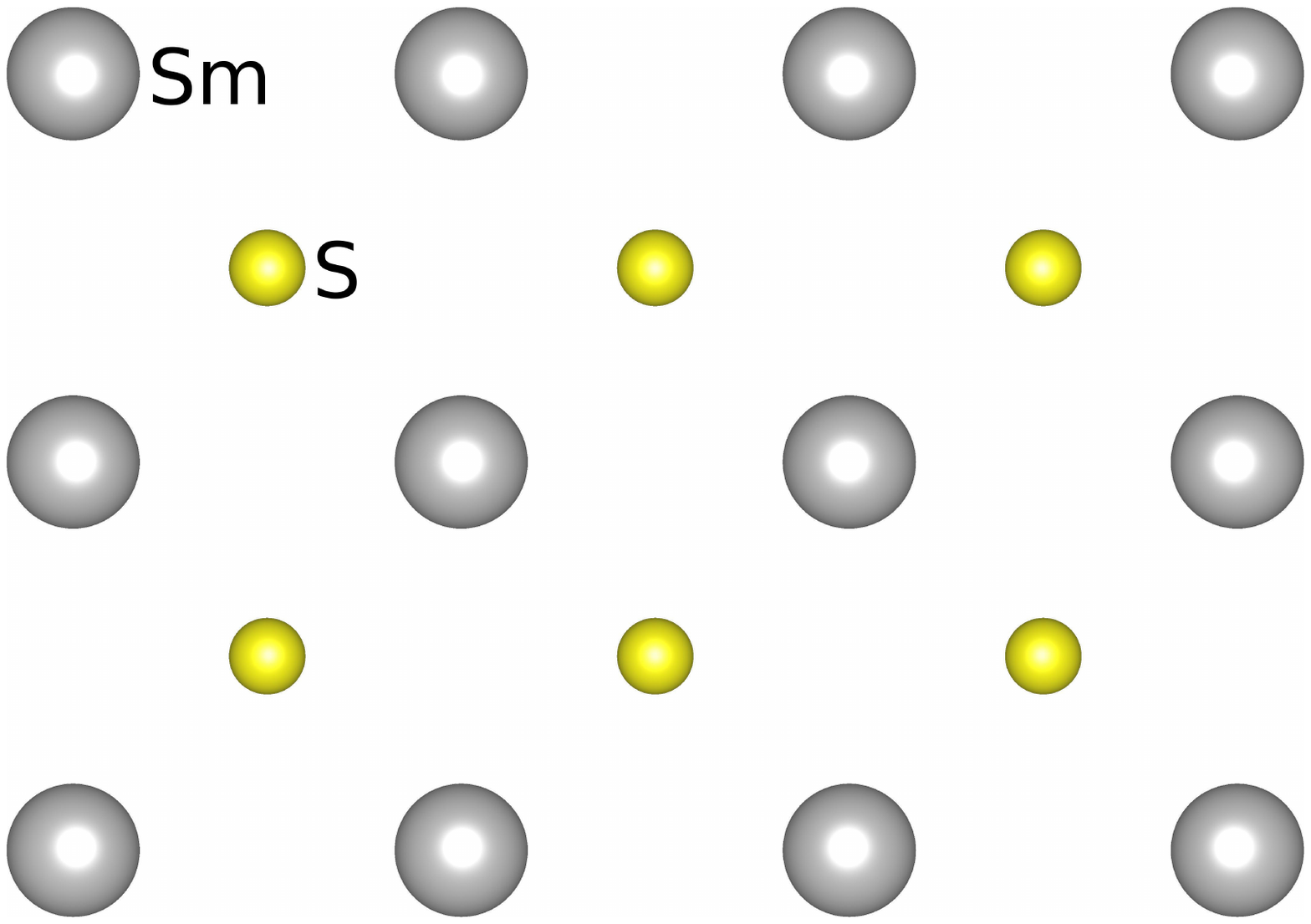}}
\caption{\label{SmS} Lattice plane of $g$-SmS.}
\end{figure}

We suggest the golden phase of SmS ($g$-SmS) as a promising candidate material. A lattice plane is shown in Fig.~(\ref{SmS}). $g$-SmS shows characteristics of a Kondo insulating state in transport, thermodynamics, and point contact spectroscopy. At temperatures low compared to this scale, the proposed lattice plane could serve as a setting to realize the frustrated $J_1-J_2$ Kondo lattice and study the anomalous Hall effect.

\end{document}